\documentclass[aps,prl,twocolumn,superscriptaddress,showpacs]{revtex4-1}
\usepackage{graphicx,amsmath,amssymb,ulem,color,braket}
\usepackage{float}
\usepackage[colorlinks]{hyperref}

\begin{document}

\title{Quantum leap: how to complete a quantum walk in a single step}

\author{Magdalena Stobi\'nska}
\email{magdalena.stobinska@gmail.com}
\thanks{Corresponding author}
\address{Institute of Theoretical Physics and Astrophysics, University of Gda\'nsk, ul.~Wita Stwosza 57, 80-952 Gda\'nsk, Poland\\\& National Quantum Information Center of Gda\'nsk, 81-824 Sopot, Poland}
\address{Institute of Physics, Polish Academy of Sciences, Al.~Lotnik\'ow 32/46, 02-668 Warsaw, Poland}
\author{Peter P. Rohde}
\address{Centre for Engineered Quantum Systems, Department of Physics and Astronomy, Macquarie University, Sydney NSW 2113, Australia}
\address{Centre for Quantum Computation and Intelligent Systems (QCIS), Faculty of Engineering \& Information Technology, University of Technology, Sydney, NSW 2007, Australia}
\author{Pawe\l{} Kurzy\'nski}
\address{Centre for Quantum Technologies, National University of Singapore, 3 Science Drive 2, 117543 Singapore, Singapore}
\address{Faculty of Physics, Adam Mickiewicz University, Umultowska 85, 61-614 Pozna\'n, Poland}

\begin{abstract}
Quantum walks provide simple models of various fundamental processes. It is pivotal to know when the dynamics underlying a walk lead to quantum advantages just by examining its statistics. A walk with many indistinguishable particles and measurements of non-classical multi-particle correlations is likely to reveal the quantum nature. The number of elements $O(n)$ in a setup realizing walks grows with their length or spread $n$. We introduce the concept of a quantum leap, a process which can be achieved with fewer or complementary resources and which in a single step simulates another long process. The process and its leap are described by the same Hamiltonian but, the latter parametrizes the evolution with a tunable parameter of a setup. In the case of walks, a leap immediately gives a probability distribution which results only after many steps. This may be appealing for simulation of processes which are lengthy or require dynamical control. We discuss a leap based on the multi-particle Hong--Ou--Mandel interference, an inherently quantum phenomenon. It reproduces a quantum walk enabling perfect state transfer through spin chains. It requires a beam splitter, two detectors and $n$ particles to mimic a walk on a chain of size $O(n)$, for time fixed by beam-splitter's reflectivity. Our results apply to a broad class of systems where the HOM-like effects can be observed, and may constitute a new approach to simulation of complex Hamiltonians with passive interferometers.
\end{abstract}

\maketitle

{\it Introduction.} Quantum optics offers precise control and manipulation of low-energy quantum light and matter. This is the first step towards efficient simulation of processes that are beyond the reach of present-day experiments and implementation of quantum-enhanced technologies. Here quantum walks (QWs) are receiving much attention since they provide simple models of various fundamental processes in nature ranging from chaos~\cite{Wojcik04}, topological phases~\cite{Kitagawa,Kitagawa2012} or photosynthesis~\cite{Mohseni,Plenio,Rebentrost,Hoyer} to universal quantum computation~\cite{Childs} and quantum search algorithms~\cite{Shenvi,Ambainis,Potocek}. Recently, numerous elementary QWs have been performed with photons~\cite{Hagai08,Schreiber10,Broome10,Peruzzo10,Schreiber11b,Matthews11,Owens11,Schreiber12,Sansoni12,Poulios14}, trapped ions~\cite{trapped1,trapped2}, neutral atoms~\cite{atoms1,atoms2,atoms3} and interacting atoms in optical lattices~\cite{lattices}.  However, verification of the quantum nature of QWs is challenging. Also the number of elements in a setup realizing QWs grows with their length or spread. These facts are unfavorable especially for long walks, where decoherence plays a detrimental role. To circumvent these limitations, an alternative way to the future technological advances in QWs implementation is asking: are there genuinely quantum processes which accelerate QWs' evolution, and which can be achieved with fewer or complementary resources? Here we examine one such example.

\begin{figure}\centering
\includegraphics[height=3cm]{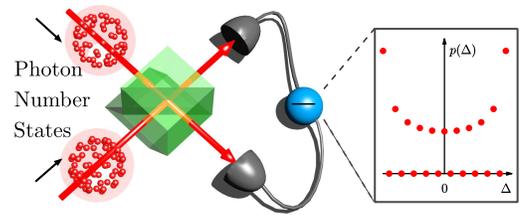}
\caption{Multiphoton Hong--Ou--Mandel interference implements a quantum leap. The probability distribution of the photon population difference $\Delta$ in the exit modes of a beam splitter simulates a continuous time quantum walk obtained after longer evolution.} 
\label{Qwalks}
\end{figure}

QWs are the quantum analogy of the classical random walk in which an initially localized particle (a walker) evolves into a superposition over many positions in space~\cite{Aharonov,Farhi1,Kempe08}. Usually, the space is discrete and can be represented as a graph. The distinctive property of a QW is U-shaped statistics of the walker's positions in a graph with variance scaling quadratically in time (ballistic spread). This is in stark contrast to a classical random walk where the probability distribution is Gaussian. However, there exist both, classical processes which mimic the QW statistics (e.g.\ classical waves explain the QW of a particle on a line~\cite{Knight,Bromberg}) and QWs characterized by the binomial distribution (e.g. QW with inhomogeneous time-dependent coin~\cite{Montero}). Thus, it is crucial to know when the dynamics underlying a walk is useful for quantum tasks, just by examining its statistics. QWs involving many indistinguishable particles are likely to reveal quantum nature, possible to verify by non-classical multi-particle correlations. 

QWs with two particles were demonstrated with photons~\cite{Sansoni12,Peruzzo10,Poulios14,White} and atoms in optical lattices~\cite{lattices}, where generalized Hong--Ou--Mandel correlations were used for testing classical inequalities. Nonclassicality of a multi-mode two-photon qutrit was inferred from selected two-mode intensity correlations. Certification of multi-partite correlations solely based on one- and two-body ones is a difficult task, and is proven only for specific qubit cases~\cite{Lewenstein}. Photonic implementation of QWs on a graph of size $O(n)$ used arrays of $O(n^2)$ beam splitters (BSs) and $O(n)$ detectors allowing $t=n$ steps~\cite{Sansoni12}.  Alternatively, arrays of $O(n)$ coupled waveguides and detectors were considered~\cite{Peruzzo10,White} however, the current technology allowed a detector behind every second waveguide. Optical lattices required $O(n)$ sites and {\it in situ} atom-resolved detection~\cite{lattices}.

Here, by means of an example, we introduce a new concept of a quantum leap (QL), a process which in a single step simulates another long process. The process and its leap are described by the same Hamiltonian but, while the former describes the evolution in time, the latter parametrizes it with a tunable parameter of a setup. For QWs, a QL immediately gives a probability distribution which results only after many steps. This may be appealing for simulation of processes which are lengthy or require dynamical control.

We discuss a QL employing the multi-particle Hong--Ou--Mandel interference~\cite{HOM87} (HOM QL), shown in Fig.~\ref{Qwalks}. This is an inherently quantum phenomenon, easily verified by the HOM interferometric contrast. The HOM QL reproduces a continuous-time QW (CTQW)~\cite{Farhi1} enabling perfect state transfer through spin chains~\cite{Christandl,Bose}. It requires a BS of reflectivity $r$, two detectors and $n$ particles to mimic a walk on a chain of size $O(n)$, for $t$ fixed by $r$. It provides a new insight into the quantum-to-classical transition~\cite{Broome10,Schreiber11b} triggered by a partial distinguishability of particles. This architecture allows control over decoherence during the evolution. 

Although our analytical results originate from quantum-photonic background, they apply to a broad class of systems and grant an intuitive access to many-body quantum dynamics.  We analyze their feasibility and conclude with a discussion and open problems.

\begin{figure*}[t]
\centering
\raisebox{3cm}{a)}\includegraphics[height=3.2cm]{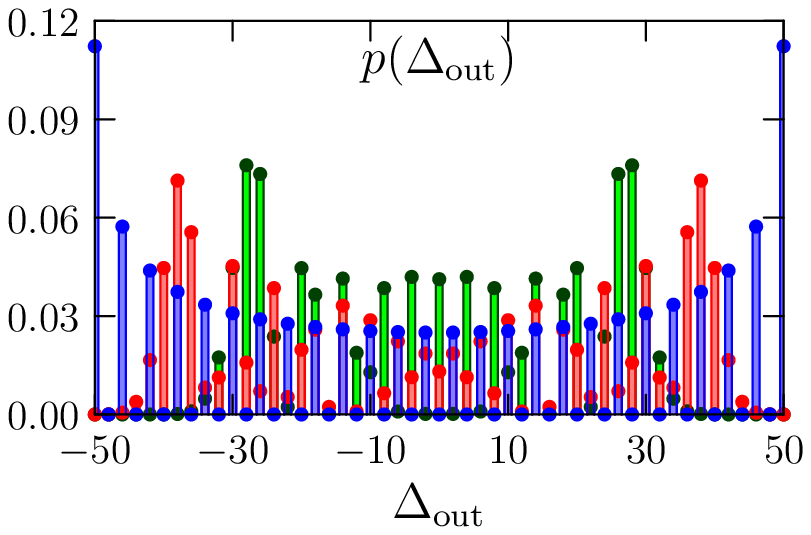}\hskip0.5cm
\raisebox{3cm}{b)}\includegraphics[height=3.2cm]{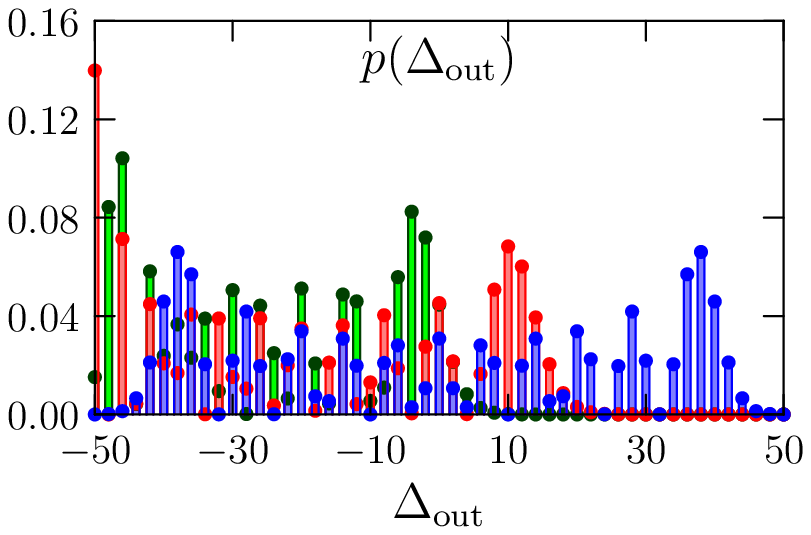}\hskip0.5cm
\raisebox{3cm}{c)}\includegraphics[height=3.2cm]{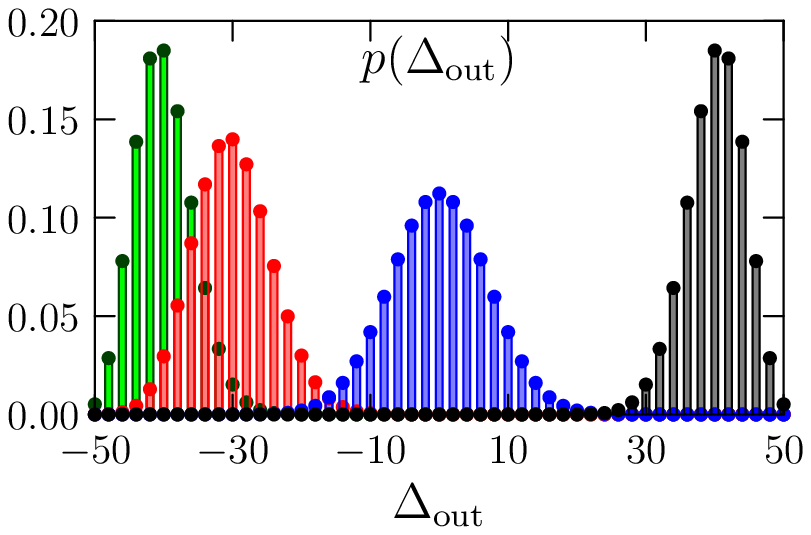}
\caption{(Color online) Probability distributions of positions of a walker in the HOM QL performed with $S=50$ photons, obtained for initial position of the walker a) $\Delta=0$, b) $-30$, c) $-50$, and reflectivity: $r =0.1$ -- green, $r =0.2$ -- red, $r=0.5$ -- blue, and $r=0.9$ -- gray.} 
\label{fig:unbalancedQW}
\end{figure*}

{\it HOM quantum leap.} The HOM interference~\cite{HOM87} is observed for two photons impinging on a BS described by the Hamiltonian 
\begin{equation}
H_{BS} = a^{\dagger}b e^{i\varphi} + ab^{\dagger}e^{-i\varphi}.
\label{bs}
\end{equation}
The annihilation operators $a$ and $b$ correspond to the interfering modes and $\varphi$ is the phase difference between the reflected and transmitted fields behind the BS. We assume $\varphi=\pi$. If the probabilities of reflection and transmission are equal, two identical photons will always leave together through the same exit port---photon bunching. Similar effects hold for multiphoton Fock-number states: photons go out only in certain configurations, e.g. such that the difference between the occupations of the exit ports is even, whereas an odd difference never occurs~\cite{Campos,MDF1,MDF2}. These interference effects, consisting of enhancing some events and canceling others, lie at the heart of the QL generated by multiphoton HOM interference.

A QW comprises at least one walker traversing a graph. A CTQW~\cite{Farhi1} takes place in the position Hilbert space where the basis states $\ket{\Delta_i}$ correspond to vertices $\Delta_i$ in the graph. In the case of a QW on a line, the position can take any integer value. The evolution of the walker is generated by a Hamiltonian $H$ which sets amplitudes of hopping from vertex $\Delta_i$ to $\Delta_j$ per unit time, $q_{i,j}=\bra{\Delta_j}H\ket{\Delta_i}$. They are zero for disjoint vertices. The walker, initially in a state $\ket{\Delta}$, after time $t$ is in a state $\ket{\Delta_{\text{out}}}=e^{-iHt}\ket{\Delta}$, where the variable $\Delta_{\text{out}}$ is governed by a doubly-peaked probability distribution $p(\Delta_{\text{out}})$ with variance scaling quadratically with time. Even in an infinitesimal time difference the walker can be transferred arbitrarily far from the starting position, albeit with low amplitude.

We will show that for interfering Fock states $\ket{S-N} \!\!=\!\! \tfrac{(a^{\dagger})^{S-N}}{\sqrt{(S-N)!}} \ket{0}$ and $\ket{N}\!\!=\!\! \tfrac{(b^{\dagger})^N}{\sqrt{N!}} \ket{0}$, $H_{BS}$ generates a CTQW on a line of length $S+1$ with specific jump amplitudes and its QL with length set by the BS reflectivity $r$.

We parametrize the input $\ket{S-N}\ket{N}$ using the mode occupation difference $\Delta = S-2N$ and denote it $\ket{\Delta}$. Behind the BS, the infinitesimal evolution turn $\ket{\Delta}$ into the superposition
\begin{equation}
H_{BS}\!\ket{\Delta} = q_{\Delta-2,\Delta}\ket{\Delta-2} + q_{\Delta+2,\Delta}\ket{\Delta+2}. 
\end{equation}
This suggests the physical interpretation of $\Delta$ as the starting position of the walker in the graph, from where it jumps to the right $\ket{\Delta+2}$ or left $\ket{\Delta-2}$. Because the length of the jump is two in the position of the particle, the evolution is confined to a region $\Delta \in \{ -S,-S+2,\dots,S-2,S\}$. The jump amplitudes 
\begin{equation}
q_{\Delta,\Delta-2}=q_{\Delta-2,\Delta}\!=\!\tfrac{1}{2}\sqrt{(S+\Delta)(S-\Delta + 2)}
\label{amplitudes}
\end{equation}
indicate an important class of CTQWs. $H_{BS}$ corresponds to the $S_x$ spin-$\tfrac{S}{2}$ matrix (in the Schwinger representation) and the two-mode Fock input state can be associated with a fictitious particle of total spin $\tfrac{S}{2}$ and eigenvalue of the $S_z$ component $\tfrac{\Delta}{2}$. Under these mappings, the HOM interference describes rotation of the $S_z$ component, equivalent to a CTQW enabling perfect state transfer in linear spin chains of length $S+1$, with inhomogeneous couplings $q_{\Delta,\Delta-2}$~\cite{Christandl}.

The HOM QL results from parameterizing the process generated by $H_{BS}\propto S_x$ with a BS reflectivity. The probability distribution of the final positions of a walker $p(\Delta_{\text{out}})$ is governed by the unitary $U_{BS}=\exp\{-i \theta H_{BS}\}$, where $r=\sin^2 \theta$ controls length of the evolution. The spectrum of $H_{BS}$ is harmonic and the evolution is periodic. For $r\!=\!1$ any input state $\ket{\Delta}$ is mapped to the opposite vertex of the graph $\ket{-\Delta}$. If $r\!=\!0$ the unitary becomes the identity and a recurrence is observed: the QW returns to the initial state. For other values of $r$, the interfering Fock states are turned into a superposition with population difference in the exit modes $\Delta_{\text{out}}$ given by  
\begin{equation}
p(\Delta_{\text{out}}) = 
\mathcal{A}
\left[
\sum_{k}  \left(\tfrac{r-1}{r} \right)^k 
\binom{f_-^{\Delta}}{k}
\binom{f_+^{\Delta}}{\tfrac{S+\Delta_{\text{out}}}{2}-k}
\right]^2,
\label{qw_distribution}
\end{equation}
where $\mathcal{A}\! =\! \tfrac{f_+^{\Delta_{\text{out}}}! f_-^{\Delta_{\text{out}}}!}{f_+^{\Delta}! f_-^{\Delta}!} (r(1-r))^S \left(\tfrac{1-r}{r} \right)^{f_+^{\Delta_{\text{out}}} - f_-^{\Delta}}$, $k$ runs from $\max\big\{0,\tfrac{\Delta_{\text{out}}-\Delta}{2}\big\}$ to $\min\big\{f_-^{\Delta},\tfrac{S+\Delta_{\text{out}}}{2}\big\}$, $f_{\pm}^x \!=\! \tfrac{S \pm x}{2}$. (For the explicit form of the output superposition and an easy-to-read form of (\ref{qw_distribution}) see the Supplementary Material). Importantly, regardless the values of $S$ and $\Delta$, the distribution (\ref{qw_distribution}) is characterized by the ballistic spread
\begin{equation}
\textrm{Var}(\Delta_{\text{out}}) \propto \tfrac{1}{4}\left(\tfrac{S^2-\Delta^2}{2} +S \right) \theta^2,
\end{equation}
adjusted by the reflectivity $r(\theta)$. (The proof is in the Supplementary Material). Thus, by fixing $r$, the multi-particle HOM interference reproduced the statistics of the discussed CTQW with the variance corresponding to an advanced stage of the evolution. This is the HOM QL.

In Fig.~\ref{fig:unbalancedQW} we plot (\ref{qw_distribution}) for various starting positions in the HOM QL performed with $S=50$ photons. In Fig.~\ref{fig:unbalancedQW}a, the walker starts in the middle of the line, i.e. $\Delta=0$, and its walk spreads uniformly in both directions, thus the distribution develops symmetrically with respect to this point. This scenario takes place if the two interfering Fock states are equal. If the initial position is shifted from the middle, $\Delta=-30$ in Fig.~\ref{fig:unbalancedQW}b, the distribution is asymmetric for all $r$ except for $r = 1/2$ and the walker tends to move in the initially preferred direction. This situation is simulated by two unequal Fock input states. The case when the walker starts in one of the most distant positions in the graph, the ends of the line $\Delta=\pm S$, correspond exactly to perfect state transfer over a chain of interacting qubits as discussed in~\cite{Christandl}. Fig.~\ref{fig:unbalancedQW}c plotted for $\Delta=-50$, shows that the initial wavepacket propagates without distortion and reaches the other end of the line with unit fidelity. This is modeled by interference of a multiphoton Fock state with the vacuum state. The odd-even comb structure is revealed in all plots: the probabilities of measuring an odd difference are zero.

{\it Quantum-to-classical transition.} In the presence of decoherence, QW statistics turn to the classical one---the binomial distribution~\cite{Kendon}. We expect this effect also for the HOM QL, where the degree of coherence is controlled by the distinguishability of interfering particles. Initially all photons carried the same polarization. Now, one beam comes in a superposition of two orthogonal polarization modes $b^{\dagger}$ and~$b^{\dagger}_{\perp}$
\begin{equation}
\label{dis}
b^{\dagger} \to \cos y \, b^{\dagger} + \sin y \, b^{\dagger}_{\perp}.
\end{equation}
$y \!\! \in \! \! (0,\tfrac{\pi}{2})$ introduces weights between the modes, and `tunes' the distinguishability. The particles are fully indistinguishable if $y\!\! =\!\! 0$, whereas if $y\!\!  =\!\!  \tfrac{\pi}{2}$ they are maximally distinguishable: $\ket{N}$ and $\ket{S-N}$ are orthogonally polarized. Transformation (\ref{dis}) leads to the interference of $\ket{N}$ with a two-mode Fock state superposition $\sum_{n=0}^{S-N} \binom{S-N}{n}^{-1/2} \cos^n y \, (\sin y)^{S-N-n} \ket{n} \ket{S-N-n}_{\perp}$, instead of the single-mode Fock state $\ket{S-N}$, as before. 

\begin{figure}\centering
\raisebox{2.4cm}{a)}\includegraphics[height=2.7cm]{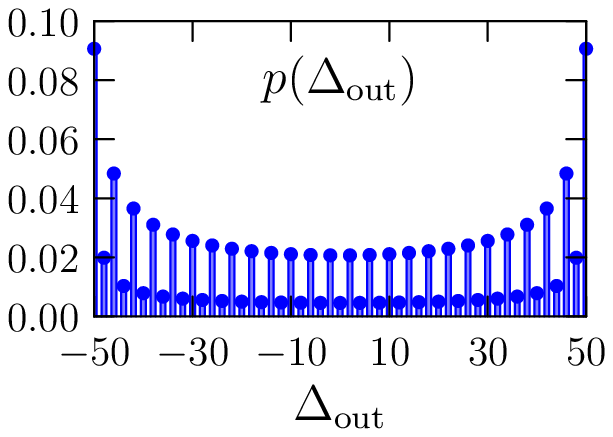}
\raisebox{2.4cm}{b)}\includegraphics[height=2.7cm]{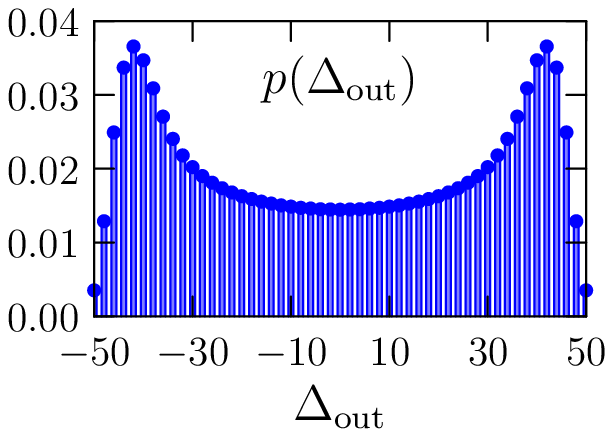}\\
\raisebox{2.4cm}{c)}\includegraphics[height=2.7cm]{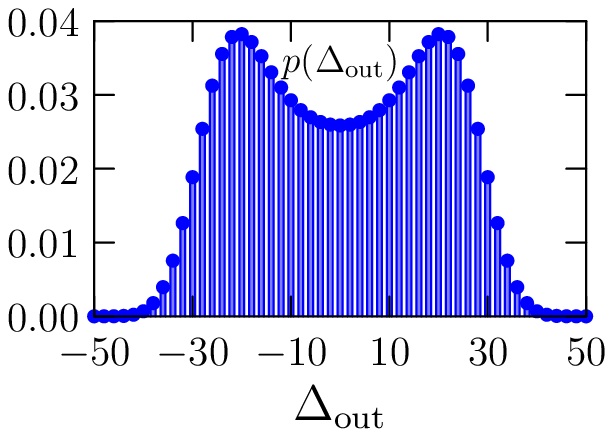}
\raisebox{2.4cm}{d)}\includegraphics[height=2.7cm]{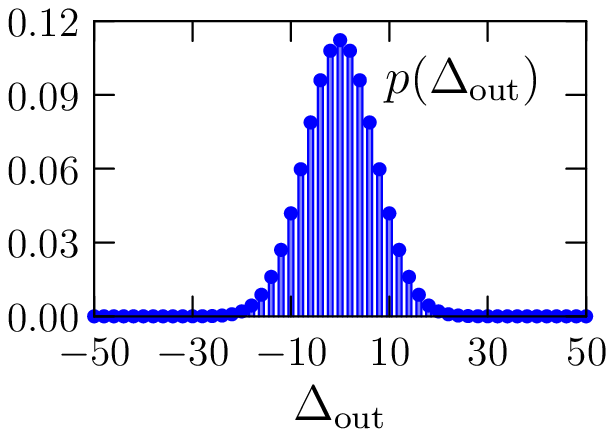}
\caption{Modification of the probability distribution depicted in Fig.~\ref{fig:unbalancedQW}a in blue ($\Delta=0$, $r = 1/2$), for various degrees of distinguishability of particles: a)~$y=\pi/24$, b)~$y=\pi/6$, c)~$y=\pi/3$ and d)~$y=\pi/2$ (fully distinguishable).}
\label{decoherence}
\end{figure}

In Fig.~\ref{decoherence} we show how the doubly-peaked distribution shown blue in Fig.~\ref{fig:unbalancedQW}a modifies if particles become distinguishable. With increasing $y$, full cancellation of certain events is impossible and the two peaks gradually shift to the center of the line and merge to create the binomial distribution for $y=\tfrac{\pi}{2}$. Distinguishability mimics decoherence because it leads to interference of the multiphoton states with the vacuum state thus, it implements the usual model describing particle loss. It causes quantum particles to behave as classical ones, whose statistics cannot mimic quantum coherence and therefore we observe a transition from the quantum to the classical domain.

{\it Multi-dimensional HOM QL.} Since a BS sees orthogonal polarizations independently, the above model generalizes to a two-dimensional case. This requires that the particle evolves independently in each direction, i.e there are no operations entangling the two polarizations. One simply considers two simultaneously interfering pairs of states: $\ket{S-N}$ and $\ket{N}$ in horizontal and $\ket{S-M}$ and $\ket{M}$ in vertical polarization. Thus now, each input beam is a two-mode beam, with fully distinguishable modes. This scenario differs from the previous decoherence model in that this time detectors are capable of distinguishing between orthogonally polarized photons. By considering orthogonal spectral modes instead of polarization, it is possible to extend the model to arbitrary dimension.

{\it Feasibility.} Various experimental platforms approach regimes of parameters necessary for demonstration of our proposal.  Atoms in optical lattices in the Mott insulator phase have fixed atom numbers per lattice site. \textit{In situ} atom-resolved detection allows measurement of second-order correlations between the sites. These systems offer tunable beam-splitter-like interactions and direct scalability to larger particle numbers~\cite{lattices}.

HOM QL was implemented with two atomic Bose-Einstein condensates (BEC) consisting of $10^4$ atoms. They were prepared in a near-single-mode mixture of Fock states using spin dynamics and a tunable beam-splitter interaction was applied~\cite{Lucke}.  Measurements were performed with resolution of 20 atoms. U-shaped statistics without the parity structure was captured. Full demonstration would require single-mode input states and high-resolution detection. Purity of Fock states (losses) was not a limiting factor, since the HOM-interference contrast is immune to losses (exemplary distributions for mixed Fock inputs and losses are shown in the Supplementary Material). 

Attempts to photonic demonstration employing bright near-single-mode squeezed vacuum twin beams (SV) are reported in~\cite{Masha}. SV were produced in parametric down conversion, followed by mode filtering. Improvement of this experiment requires the creation of Fock states. They could be conditionally prepared from two-mode SV states, which posses perfect photon-number correlations between the modes (the signal and idler). The idler would be measured by a photon counting detector, whereas the signal would end up in a known Fock state. For the HOM QL, one would need two such sources and detectors. The range of $S=10$ would match best superconducting transition-edge sensors (TESs)~\cite{TES1} with photon-counting efficiencies near $100\%$ and well-resolved photon-number peaks up to around ten photons~\cite{TES2}. A BS of a controllable reflectivity can be implemented by a Mach-Zehnder interferometer with a dynamically controlled phase shift applied to one of the inputs~\cite{Zeilinger2011,OBrien2012}. (Fig.~\ref{fig:unbalancedQW} for $S=10$ is shown in the Supplementary Material).

{\it Conclusions.} Quantum leaps may be an important concept in the framework of quantum simulations. They quickly restore long evolution by mapping physical time to a controllable parameter of a setup while preserving an original Hamiltonian. This is appealing for simulation of processes requiring dynamical control, e.g. transport in presence of dynamical disorder, using passive setups. 

The HOM interference of many indistinguishable particles provides a link between a walk of a quantum nature (useful for quantum tasks) and a QL. It is characterized by the second-order visibility, which for two identical Fock states impinging on a symmetric BS is quantum $v^{(2)}=1/(2-1/n)>1/2$ and cannot be explained within classical theory (the case of two non-identical Fock states is discussed in the Supplementary Material). Moreover, it imparts an effective nonlinearity leading to the exchange interaction between photons at a BS~\cite{Peter}. Up to date, nonlinear effects and interacting bosons have not been explored in photonic quantum walks. Importance of generalized HOM effects for practical purposes has been recognized~\cite{Crespi}. Combining the HOM interference with a standard QW architecture will constitute setup enabling observation of large-scale interference, boson sampling~\cite{Gard,Scheel}, multi-particle entanglement and non-linear phenomena~\cite{Peter}. 

The HOM QL reproduces the statistics of a CTQW allowing perfect state transfer over spin chains. Its variance is set by reflectivity of the HOM BS and reveals ballistic spread. It replicates quantum-to-classical transition triggered by partial distinguishability of particles and establishes a link between the fundamental indistinguishability of quantum particles and the wavelike coherent nature of the walk. We have discussed one-dimensional HOM QL but by exploiting various degrees of freedom, the scheme may be generalized to higher dimensions. 

Implementations of QWs without interferometers are gaining a lot of interest now~\cite{Cardano}. Our approach represents a substantial saving in elements and detectors -- we reduce the complexity of the system to having a single BS, two detectors and $O(n)$ particles to implement a CTQW with $n$ positions. This allows control of decoherence during the evolution, by shifting it to the initial state preparation and detection scheme. 

We have discussed the feasibility of HOM QLs with optical lattices, BECs and optical systems.  Our insights are of a general relevance and apply to a broad class of bosons: diffracting neutron~\cite{Neutron} and X rays~\cite{Xrays}, surface plasmon polaritons~\cite{Plasmon}, and fermions~\cite{fermions1,fermions2} where the HOM-like effects were observed.

Our results lead to open problems. Is it possible to find interferometers giving QWs with arbitrary jump amplitudes? One can consider e.g. QWs with interacting BECs or atoms in optical lattices with larger particle numbers per lattice site, leading to generalization of the results presented in~\cite{Lucke} and~\cite{lattices}. How does the interaction influence the dynamics of the walk and the observed probability distribution? What classes of Hamiltonians can be simulated with passive interferometers, perhaps including complex input states? In this case, a single step of a QL would be understood as a direct mapping of the input to the output state. The intermediate operations implemented in such setup would not necessarily correspond to successive steps of evolution.

We believe that this work will establish a new approach to quantum simulations and to bottom-up study of the dynamics of complex systems.

{\it Acknowledgments.} MS and PK thank Anton Zeilinger and Fabio Sciarrino for fruitful discussions. MS was supported by the EU 7FP Marie Curie Career Integration Grant No. 322150 ``QCAT'', NCN grant No. 2012/04/M/ST2/00789, MNiSW co-financed international project No. 2586/7.PR/2012/2 and MNiSW Iuventus Plus project No. IP 2014 044873. PPR was supported by the Australian Research Council Centre of Excellence for Engineered Quantum Systems (Project number CE110001013), and acknowledges financial support from Lockheed Martin. PK is supported by the National Research Foundation and the Ministry of Education in Singapore.

\section*{Appendix A: HOM QL probability distribution $p(\Delta_{\text{out}})$}

We will now consider an unbalanced beam splitter with reflectivity $r$, which transforms the input annihilation operators $a$ and $b$ in the following way: $a_r = \sqrt{1-r} a + \sqrt{r} b$, $a_t = -\sqrt{r} a + \sqrt{1-r} b$. Note that $r$ is the single-photon reflectivity because it gives the probability of reflection of a photon in the case if a single photon impinges on the beam splitter. The incoming Fock state $\ket{K}_a\ket{L}_b$, where $\ket{K}=\tfrac{(a^{\dagger})^K}{\sqrt{K!}} \ket{0}$ and $\ket{L}=\tfrac{(b^{\dagger})^L}{\sqrt{L!}} \ket{0}$, is turned into the following superposition
\begin{align}
\ket{\psi_\mathrm{out}} =& \tfrac{1}{\sqrt{K! L!}} (\sqrt{1-r} a_r^\dag - \sqrt{r} a_t^\dag)^K (\sqrt{r} a_r^\dag + \sqrt{1-r} a_t^\dag)^L \ket{0} \nonumber\\
=& \tfrac{1}{\sqrt{K! L!}} \sum_{k=0}^K \sum_{l=0}^L \binom{K}{k} \binom{L}{l} (-1)^{K-k} \nonumber\\
{}&\sqrt{r}^{K-k+l} \sqrt{1-r}^{L-l+k} {a_r^\dag}^{k+l} {a_t^\dag}^{K-k+L-l} \ket{0}.
\end{align}
We are interested in the probability distribution of measurement outcomes of photon counting detectors located behind the beam splitter $|\langle p,q|\psi_\mathrm{out}\rangle|^2$. Two Fock states $\ket{K}$ and $\ket{L}$ can always be parametrized by the total photon number $S=K+L$ and the population difference $\Delta = K-L$. In this case $\ket{K} = \ket{\tfrac{S+\Delta}{2}}$ and $\ket{L}= \ket{\tfrac{S-\Delta}{2}}$. Since $p+q=S$ as well, and $p-q=\Delta_{\text{out}}$ the probability distribution reads
\begin{equation}
\begin{aligned}[t]
&p(\Delta_{\text{out}}) =
\dfrac{\left(\tfrac{S-\Delta_{\text{out}}}{2}\right)!\left(\tfrac{S+\Delta_{\text{out}}}{2}\right)!}{\left(\tfrac{S-\Delta}{2}\right)!\left(\tfrac{S+\Delta}{2}\right)!} 
\big(r(1-r)\big)^S \left(\tfrac{1-r}{r} \right)^{\tfrac{\Delta-\Delta_{\text{out}}}{2}} 
\times{}\\&\quad
\left[\rule{0cm}{1cm}\right.
\sum_{k=\max\left(0,\tfrac{\Delta_{\text{out}}-\Delta}{2}\right)}^{\min\left(\tfrac{S-\Delta}{2},\tfrac{S+\Delta_{\text{out}}}{2}\right)}
\!\!\!
\binom{\tfrac{S-\Delta}{2}}{k}
\binom{\tfrac{S+\Delta}{2}}{\tfrac{S+\Delta_{\text{out}}}{2}-k}
\left(\tfrac{r-1}{r} \right)^k
\left.\rule{0cm}{1cm}\right]^2\!\!\!.
\end{aligned}
\end{equation}
The above distribution corresponds to Eq.~(4) in the main text. It is shown for $S=10$ in Fig.~\ref{fig:unbalancedQWS10} below.  

\begin{figure*}\centering
\raisebox{3.7cm}{a)}\includegraphics[height=4cm]{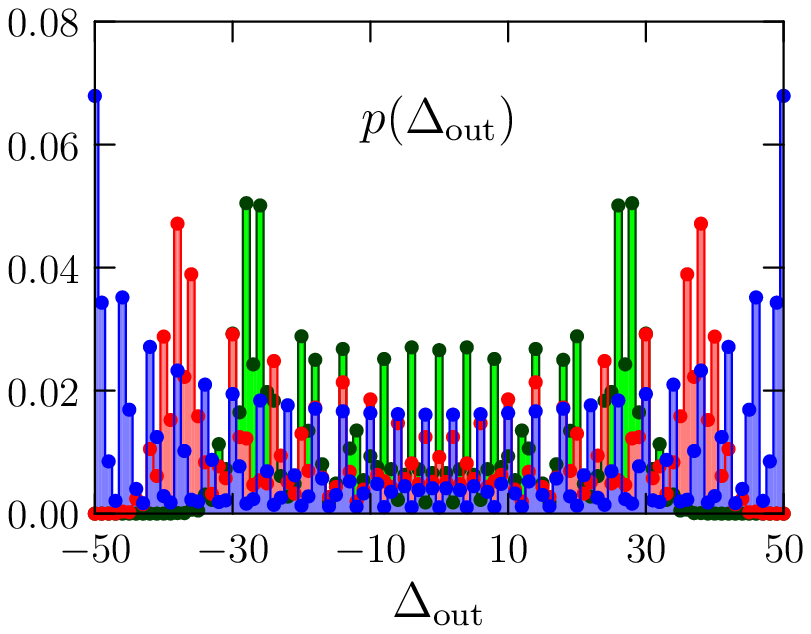}
\raisebox{3.7cm}{b)}\includegraphics[height=4cm]{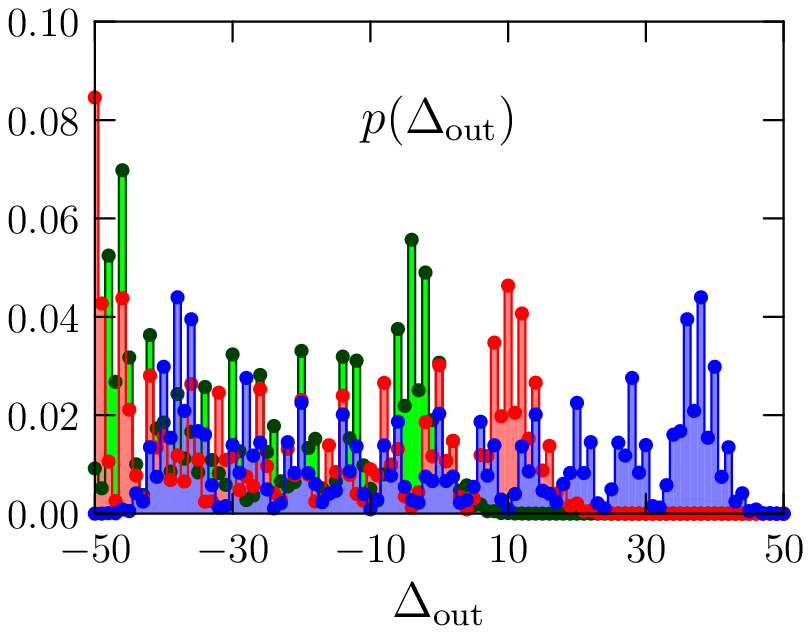}
\raisebox{3.7cm}{c)}\includegraphics[height=4cm]{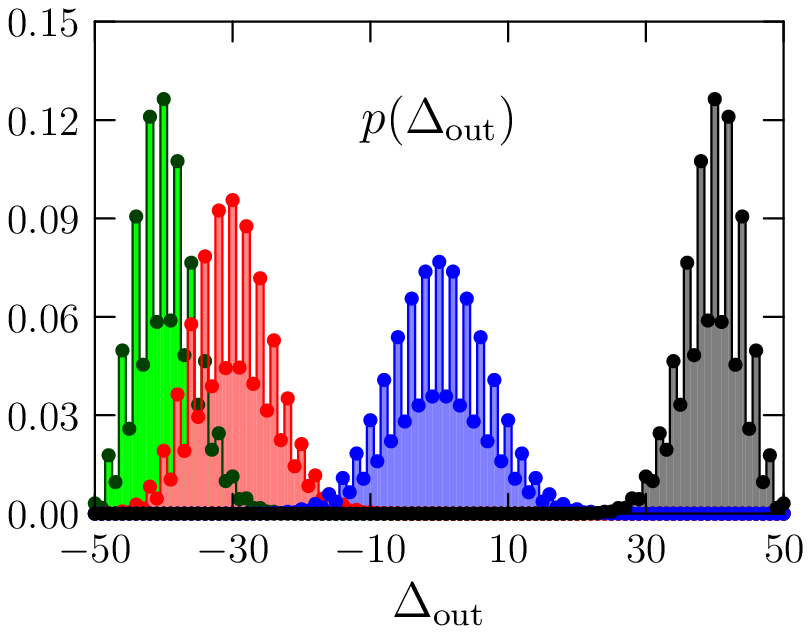}
\caption{(Color online) Probability distributions from Fig.~\ref{fig:unbalancedQW} in the main text computed for input Fock states of purity $0.47$, $S=50$ a) $\Delta=0$, b) $-30$, c) $-50$.}
\label{purity}
\end{figure*}

In a similar way, one can compute the probability distribution for mixed Fock input states $\rho_K = \sum_{k=0}^K \binom{K}{k}\eta^{K-k}(1-\eta)^k |k\rangle\langle k|$ and $\rho_L = \sum_{l=0}^L \binom{L}{l}\eta^{L-l}(1-\eta)^l |l\rangle\langle l|$, with purity equal $\mathrm{Tr}\rho^2$, and lossy detectors.
Fig.~\ref{purityS10} displays the distributions from Fig.~\ref{fig:unbalancedQWS10} computed for input Fock states with purity $0.83$, whereas Fig.~\ref{purity} shows the distributions from Fig.~\ref{fig:unbalancedQW} in the main text computed for input Fock states with purity $0.47$. Fig.~\ref{fig:losses} shows an array of the distributions obtained for $S=10$ and various purities of input Fock states and amount of losses.

\begin{figure*}\centering
\raisebox{3.7cm}{a)}\includegraphics[height=4cm]{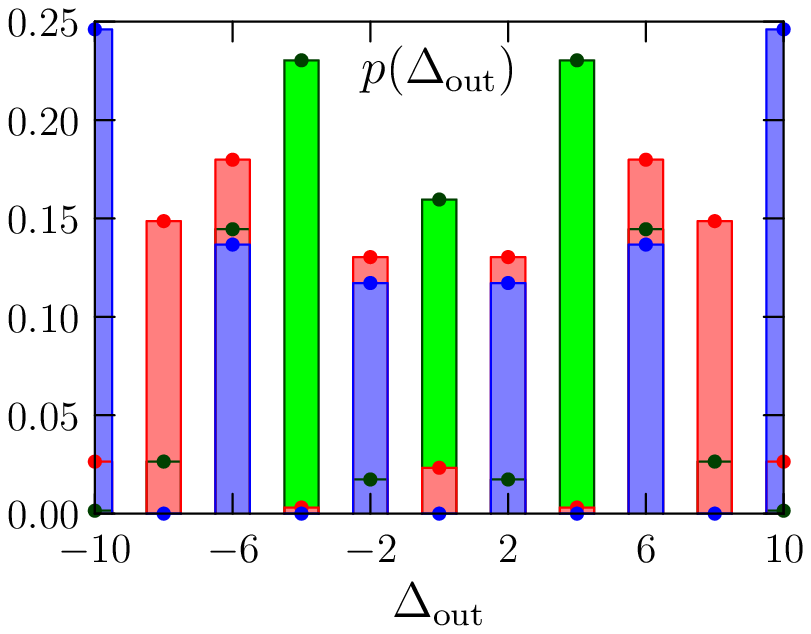}
\raisebox{3.7cm}{b)}\includegraphics[height=4cm]{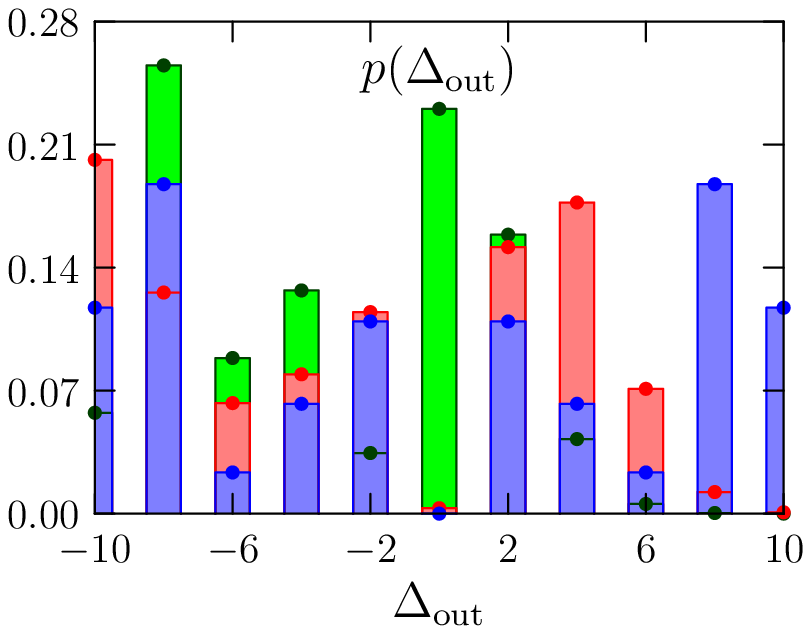}
\raisebox{3.7cm}{c)}\includegraphics[height=4cm]{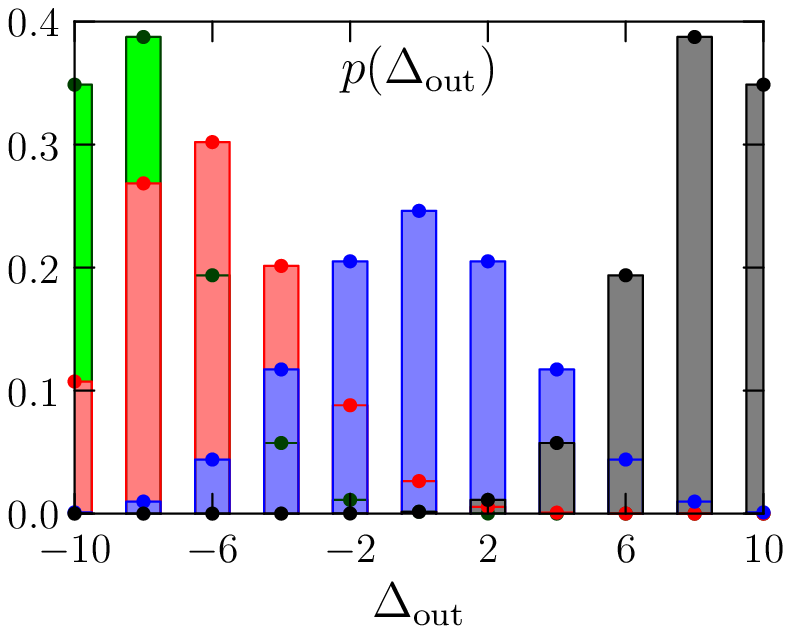}
\caption{(Color online) Probability distributions of positions of a walker in the HOM QL performed with $S=10$ photons, obtained for initial position of the walker a) $\Delta=0$, b) $-4$, c) $-10$ and reflectivity: $r=0.1$ -- green, $r=0.2$ -- red, $r=0.5$ -- blue, and $r=0.9$ -- gray.} 
\label{fig:unbalancedQWS10}
\end{figure*}

\begin{figure*}\centering
\raisebox{3.7cm}{a)}\includegraphics[height=4cm]{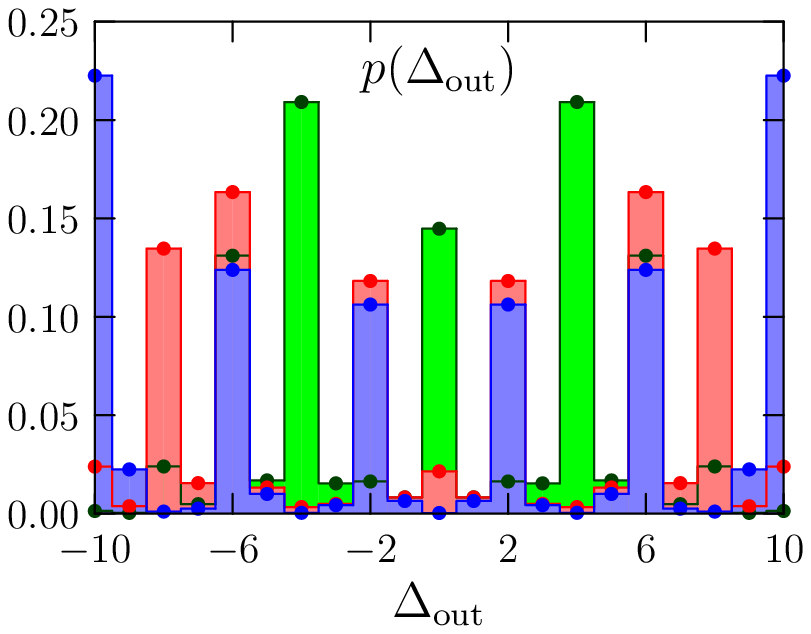}
\raisebox{3.7cm}{b)}\includegraphics[height=4cm]{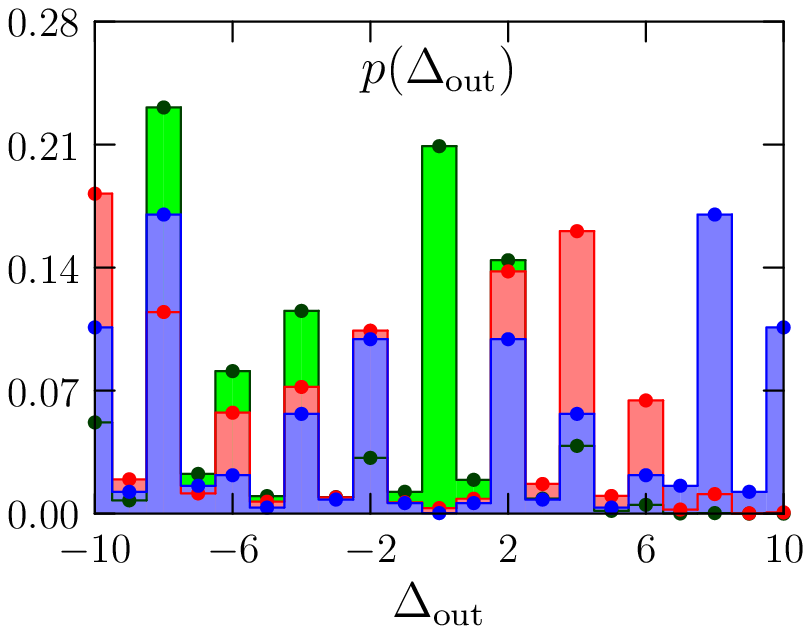}
\raisebox{3.7cm}{c)}\includegraphics[height=4cm]{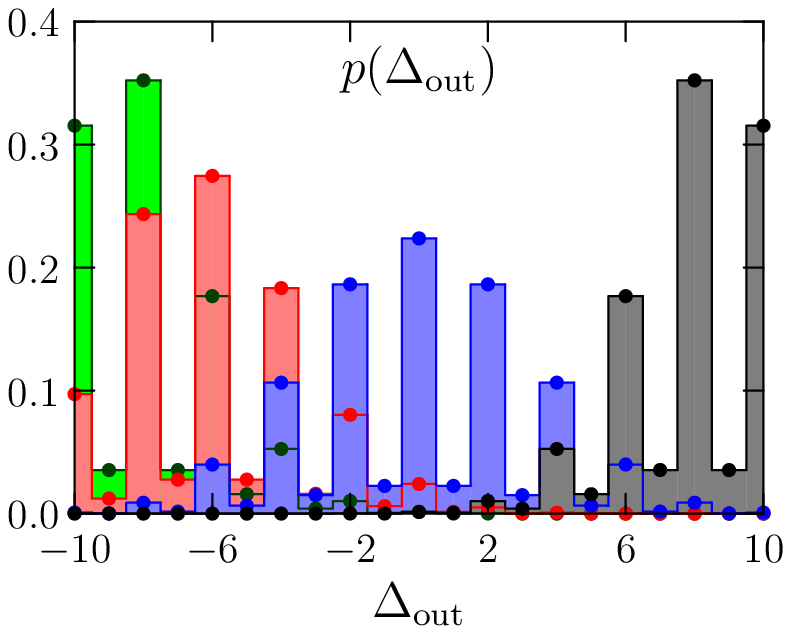}
\caption{(Color online) Probability distributions from Fig.~\ref{fig:unbalancedQWS10} computed for input Fock states of purity $0.83$, $S=10$ a) $\Delta=0$, b) $-4$, c) $-10$.}
\label{purityS10}
\end{figure*}

\section*{Appendix B: Ballistic spread of $p(\Delta_{\text{out}})$}

At first let us note that the beam splitter Hamiltonian 
\begin{equation}
H_{BS} = a^{\dagger}b e^{i\varphi} + ab^{\dagger}e^{-i\varphi}
\label{bsAppendix}
\end{equation}
(Eq.~(1) in the main text), for $\varphi=\pi$, corresponds to the Schwinger representation of the $S_x$ spin-$\tfrac{S}{2}$ matrix, defined as $S_x = \tfrac{a^{\dagger}b + ab^{\dagger}}{2}$. Secondly, a two-mode Fock state $\ket{S-N}\ket{N}$ can be associated with a fictitious particle of total spin $\tfrac{S}{2}$ and eigenvalue of the $S_z$ component equal to $\tfrac{\Delta}{2}$, where $S_z = \tfrac{a^{\dagger}a - b^{\dagger}b}{2}$. Now $\ket{S-N}\ket{N} \equiv \ket{\Delta}$. Here $\Delta= S-2N$ denotes the initial mode occupation difference. The evolution generated by the beam splitter's Hamiltonian $H_{BS} \propto S_x$ is parametrized by the beam-splitter's reflectivity $r$, $U_{BS}=\exp\{-i \theta H_{BS}\}$ where $r=\sin^2 \theta$. 

Under these mappings, the HOM QL is in one-to-one correspondence with a spin-$\tfrac{S}{2}$ particle whose evolution is given by the Hamiltonian proportional to the $S_x$ operator. In this picture the position operator (of the walker in the graph) is proportional to the $S_z$ operator and the states with well defined position correspond to the eigenstates of $S_z$: $|\tfrac{S}{2},\tfrac{\Delta}{2}\rangle$. It is therefore straightforward to show that in the Heisenberg picture the $S_z$ evolves as
\begin{equation}
S_z(\theta) = S_z \cos \theta + S_y \sin \theta.
\end{equation}
Note that $S_y$ is a linear combination of rising $S_+=a^{\dagger}b$ and lowering $S_- = b^{\dagger}a$ operators, $S_y = \frac{i(S_- - S_+)}{2}$, therefore for the states $|\tfrac{S}{2},\tfrac{\Delta}{2}\rangle$ the average value of $S_y$ is zero and we obtain 
\begin{equation}
\langle S_z(\theta) \rangle = \langle S_z \rangle \cos \theta.
\end{equation}
Moreover,
\begin{equation}
S_z^2(\theta) = S_z^2 \cos^2 \theta + S^2_y \sin^2 \theta + \{S_y,S_z\} \cos \theta \sin \theta.
\end{equation}
Due to the same reason as above, for the states $|\tfrac{S}{2},\tfrac{\Delta}{2}\rangle$ the average value of $S_z^2(\theta)$ equals
\begin{equation}
\langle S_z^2(\theta) \rangle = \langle S_z^2 \rangle \cos^2 \theta + \langle S^2_y \rangle \sin^2 \theta.
\end{equation}

\begin{figure}[h]\centering
\raisebox{2.4cm}{a)}\includegraphics[height=3.1cm]{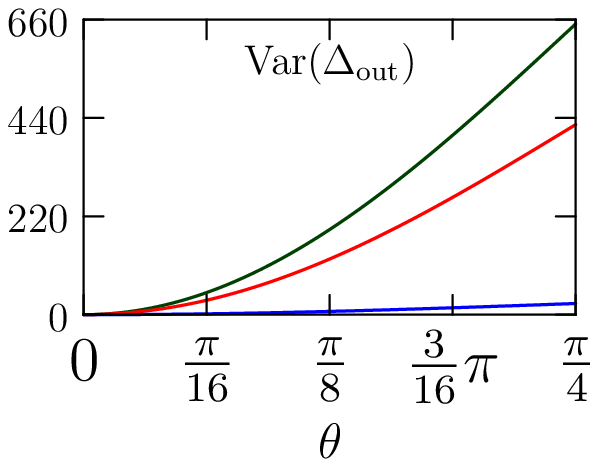}
\raisebox{2.4cm}{b)}\includegraphics[height=3.1cm]{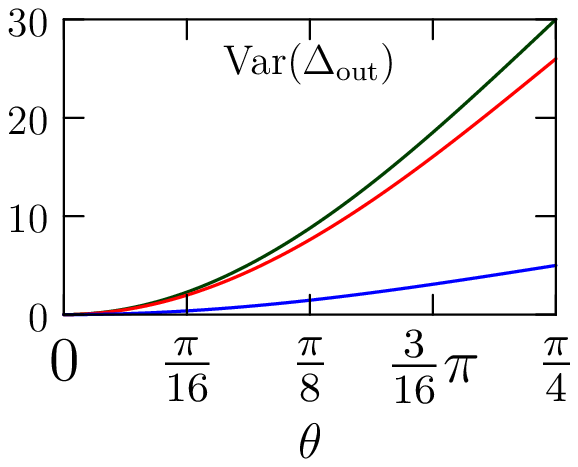}\\
\raisebox{2.4cm}{c)}\includegraphics[height=3.1cm]{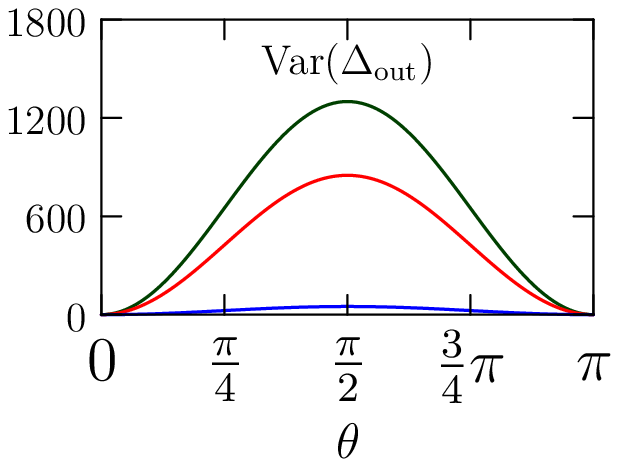}
\raisebox{2.4cm}{d)}\includegraphics[height=3.1cm]{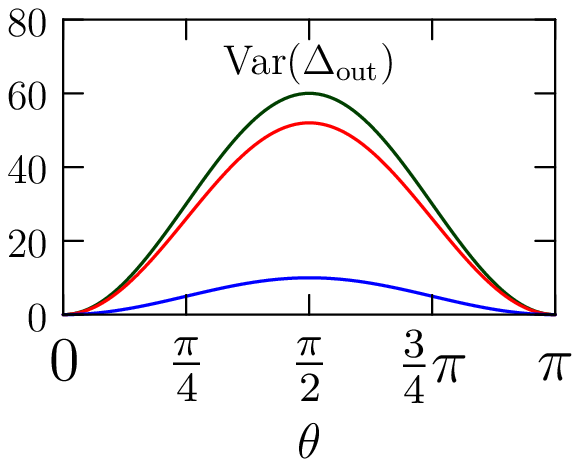}
\caption{Variance of the probability distributions of final positions of a walker in the HOM QL for a) \& c) $S=50$ and the initial position of the walker $\Delta=0$ -- green, $\Delta=-30$ -- red and $\Delta=-50$ -- blue b) \& d) $S=10$ and $\Delta=0$ -- green, $\Delta=-4$ -- red and $\Delta=-10$ -- blue.}
\label{variances}
\end{figure}

In addition, since $|\tfrac{S}{2},\tfrac{\Delta}{2}\rangle$ are eigenstates of $S_z$, we have $\langle S_z^2 \rangle = \langle S_z \rangle^2$. The value $\langle S_y^2 \rangle$ can be evaluated in the following way. At first, we note that
\begin{equation}
S_y^2= \frac{-(S_- - S_+)^2}{4} = \frac{-S_-^2 - S_+^2 + S_+ S_- + S_- S_+}{4}.
\end{equation}
For the states $|\tfrac{S}{2},\tfrac{\Delta}{2}\rangle$ we have
\begin{equation}
\langle S_y^2 \rangle = \frac{\langle S_+ S_- \rangle  + \langle S_- S_+ \rangle }{4} = \frac{\tfrac{S}{2}(\tfrac{S}{2}+1) - \left(\tfrac{\Delta}{2}\right)^2}{2}.
\end{equation}
Therefore
\begin{equation}
\textrm{Var}[S_z(\theta)] = \langle S_z^2(\theta) \rangle - \langle S_z(\theta) \rangle^2 = \frac{\tfrac{S}{2}(\tfrac{S}{2}+1) - \left(\tfrac{\Delta}{2}\right)^2}{2} \sin^2 \theta.
\end{equation}
The variance oscillates because the HOM QL takes place on a finite graph. However, for short times the boundary effects cannot play any significant role. Indeed, by approximating $\sin \theta \approx \theta $ we get

\begin{equation}
  \textrm{Var}[S_z(\theta)] \approx \tfrac{1}{4} \left( \tfrac{S^2 - \Delta^2}{2} + S \right) \theta^2,
\end{equation}

\noindent
which confirms ballistic spreading. In the main text we denote $\textrm{Var}[S_z(\theta)]$ by $\textrm{Var}(\Delta_{\text{out}})$. Fig.~\ref{variances} shows the variance of the probability distributions of final positions of a walker in the HOM QL as a function of the time of the walk, for various values of $S$ and $\Delta$.

\section*{Appendix C: Second-order visibility in the HOM interference of Fock states}

\begin{figure}[h]\centering
\raisebox{2.7cm}{a)}\includegraphics[height=3cm]{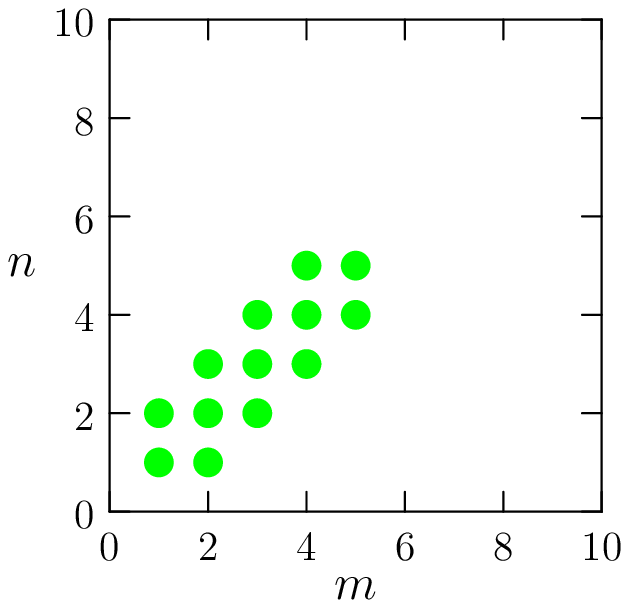}
\raisebox{2.7cm}{b)}\includegraphics[height=3cm]{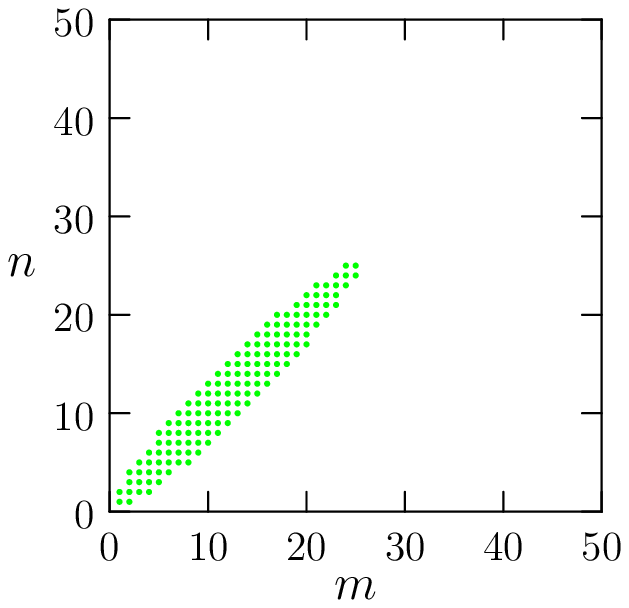}\\
\raisebox{2.7cm}{c)}\includegraphics[height=3cm]{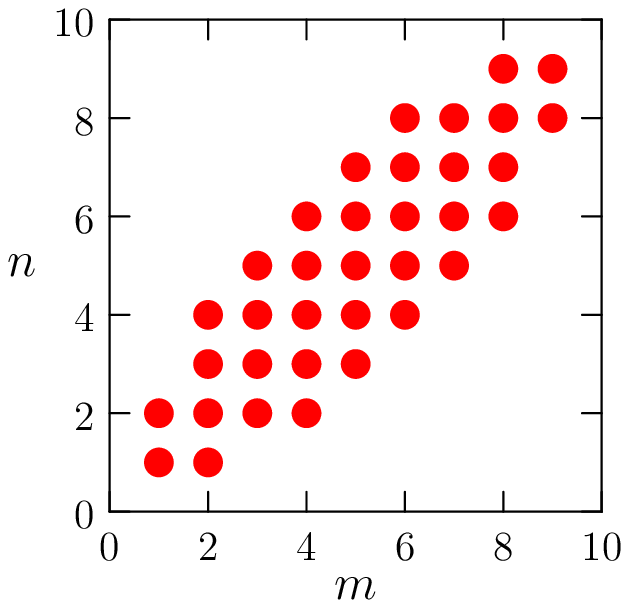}
\raisebox{2.7cm}{d)}\includegraphics[height=3cm]{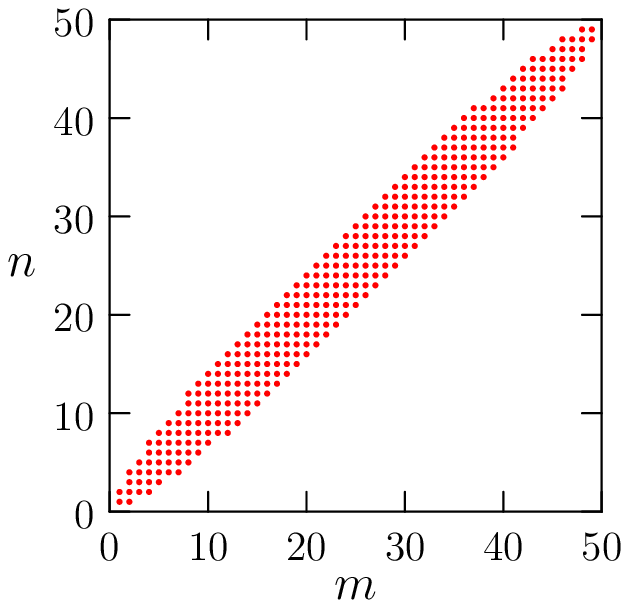}\\
\raisebox{2.7cm}{e)}\includegraphics[height=3cm]{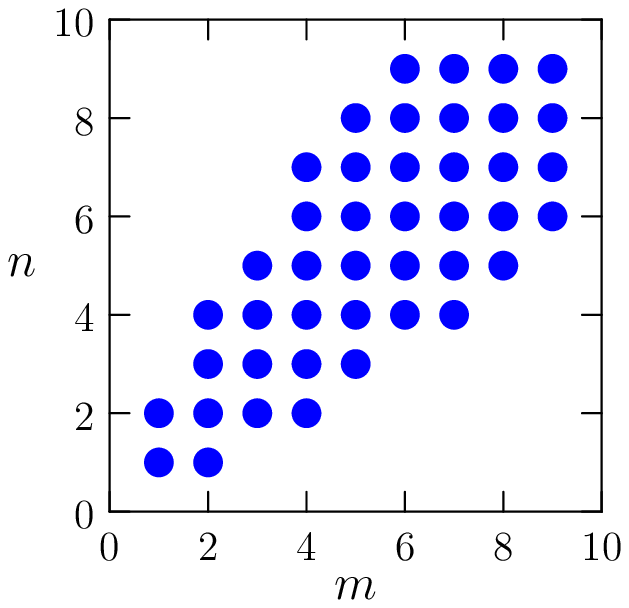}
\raisebox{2.7cm}{f)}\includegraphics[height=3cm]{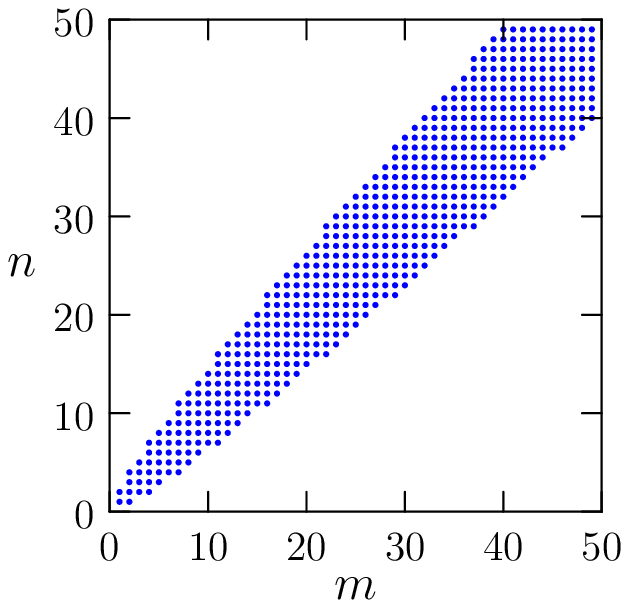}
\caption{The range where the second-order visibility of the HOM interference $v^{(2)}$ exceeds $1/2$ for two non-identical Fock states $\lvert n\rangle$ and $\lvert m\rangle$, computed for $n,m\le 10$ (left column) and $n,m\le 50$ (right column) and a) $r=0.36$, b) $r=0.43$, c) $r=0.39$, d) $r=0.45$, e) \& f) $r=0.5$.}
\label{fig:visibility}
\end{figure}

Visibility of the HOM interference at a beam splitter with reflectivity $r$ and transmitivity $t=1-r$ is given by 
\begin{eqnarray}
v^{(2)} &= \dfrac{2 rt\, \langle : n_a n_b:\rangle}{rt\, \langle : n_a^2 + n_b^2:\rangle + (r^2+t^2)\, \langle : n_a n_b:\rangle},
\end{eqnarray}
where $n_a = a^{\dagger}a$, $n_b = b^{\dagger}b$, the annihilation operators $a$ and $b$ correspond to the interfering modes and $:\, :$ denotes the normal ordering. If $v^{(2)}>1/2$, the HOM interference does not have explanation within classical theory. Importantly, the visibility is immune to losses. Thus, its value is the same for two interfering pure and mixed Fock states and can be verified with lossy detectors.  For two non-identical Fock states $\lvert n\rangle$ and $\lvert m\rangle$ it equals
\begin{eqnarray}
v^{(2)} &= \dfrac{2r t\, nm}{rt\,(n^2-n+m^2-m)+(r^2+t^2)\,nm}.
\end{eqnarray}
 Fig.~\ref{fig:visibility} shows the range of $m$ and $n$ for which the visibility exceeds $1/2$ thus, takes quantum values.

\begin{figure*}[p]\centering
\begin{tabular}{llcccc}
&&\multicolumn{4}{c}{Purity}\\
& Losses& $0.21$& $0.41$& $0.83$& $1.00$\\
\raisebox{2cm}{a)}&
\raisebox{1.4cm}{$0\%$}&
\includegraphics[height=2cm]{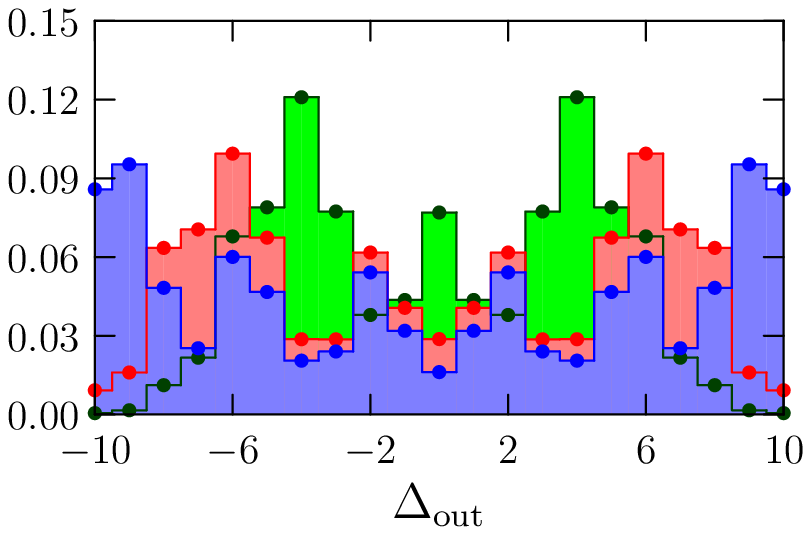}&
\includegraphics[height=2cm]{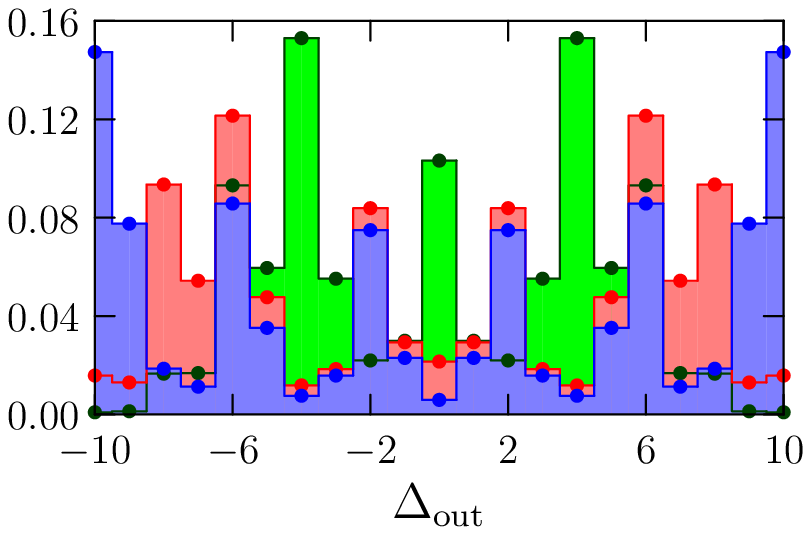}&
\includegraphics[height=2cm]{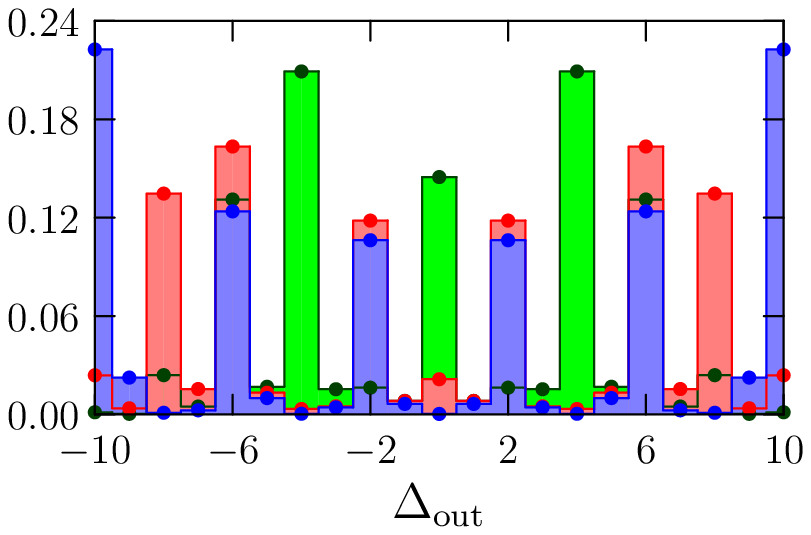}&
\includegraphics[height=2cm]{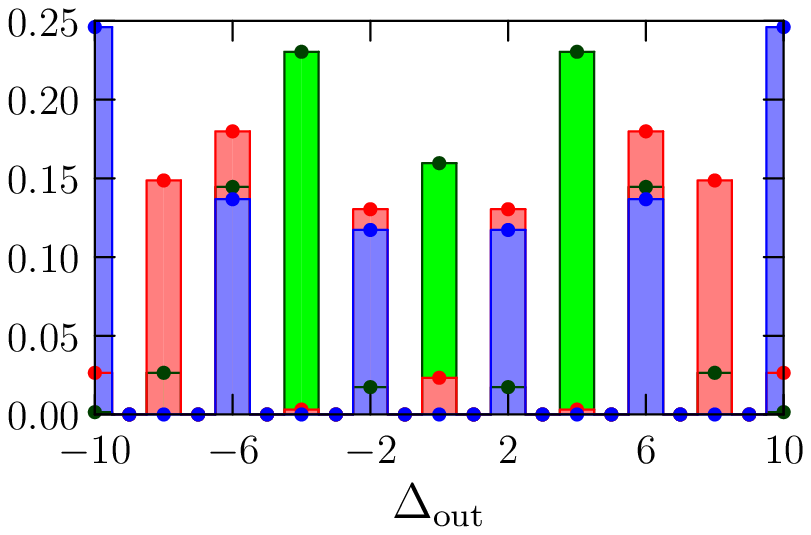}\\
&\raisebox{1.4cm}{$10\%$}&
\includegraphics[height=2cm]{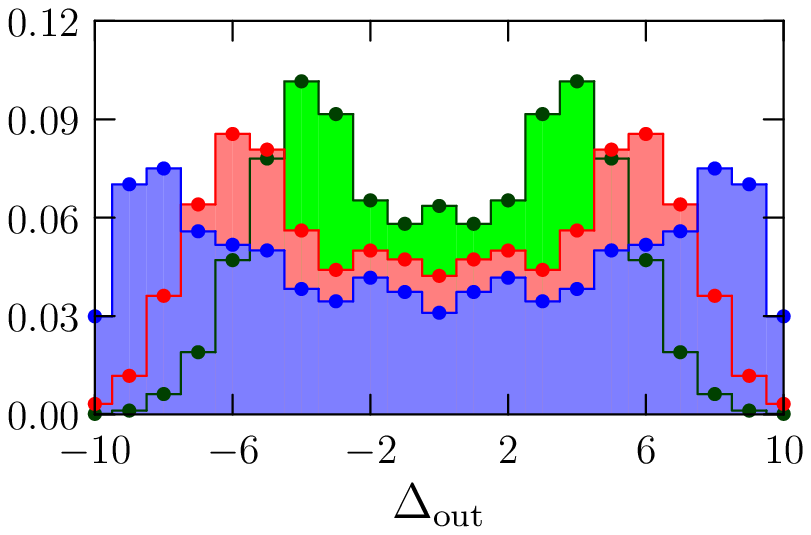}&
\includegraphics[height=2cm]{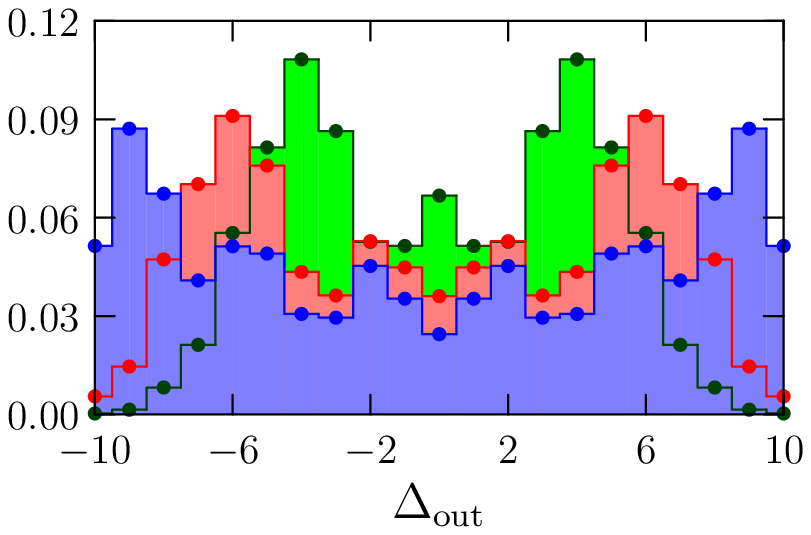}&
\includegraphics[height=2cm]{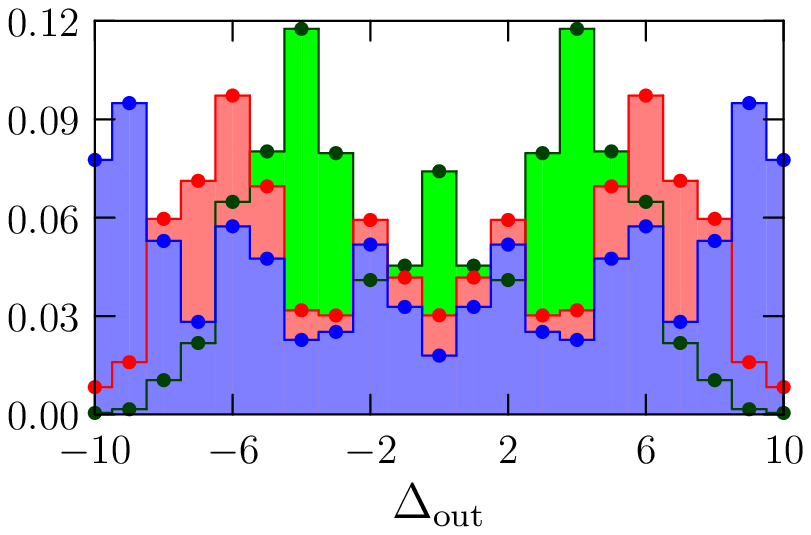}&
\includegraphics[height=2cm]{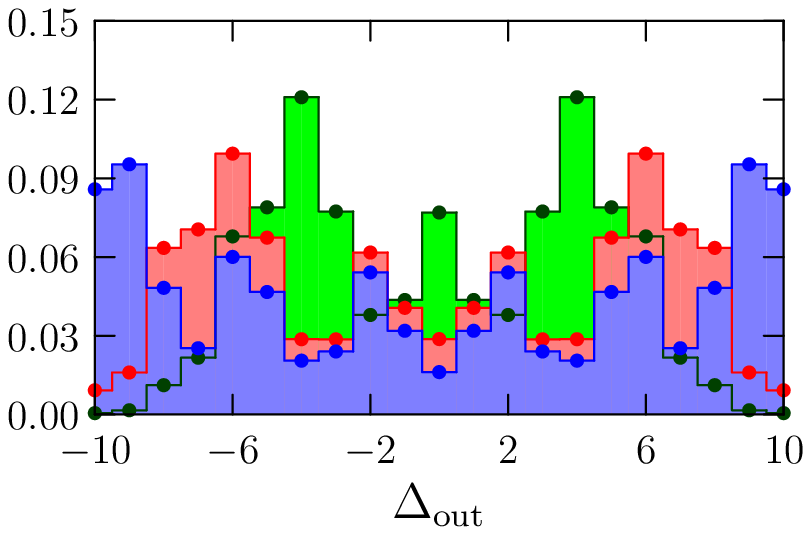}\\
&\raisebox{1.4cm}{$20\%$}&
\includegraphics[height=2cm]{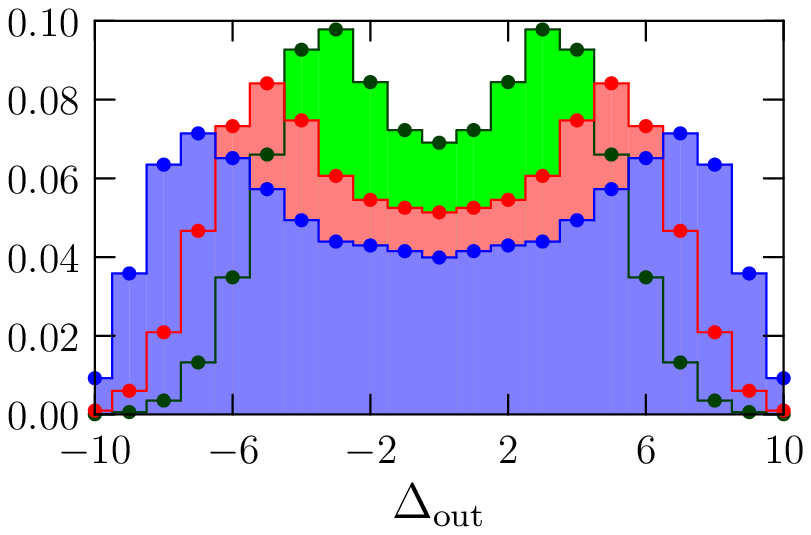}&
\includegraphics[height=2cm]{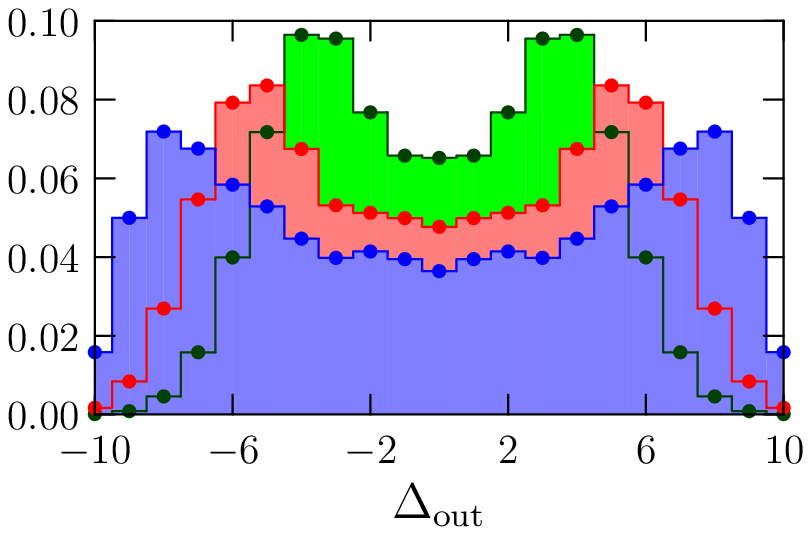}&
\includegraphics[height=2cm]{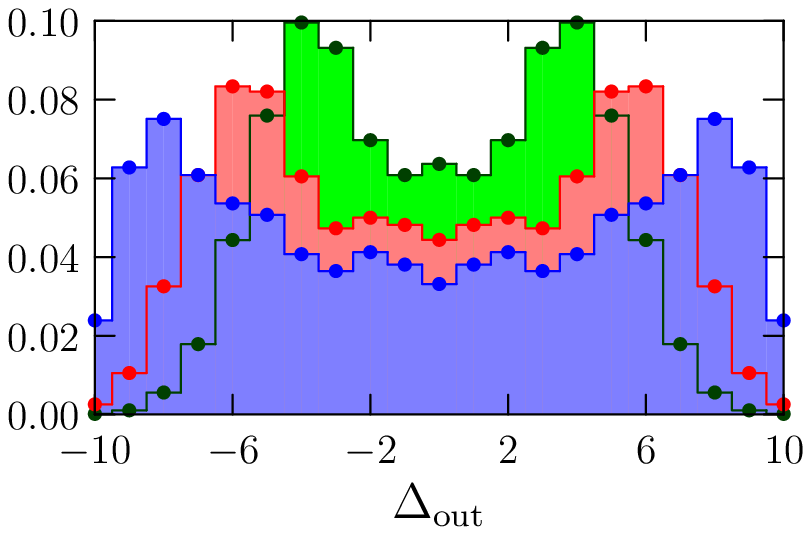}&
\includegraphics[height=2cm]{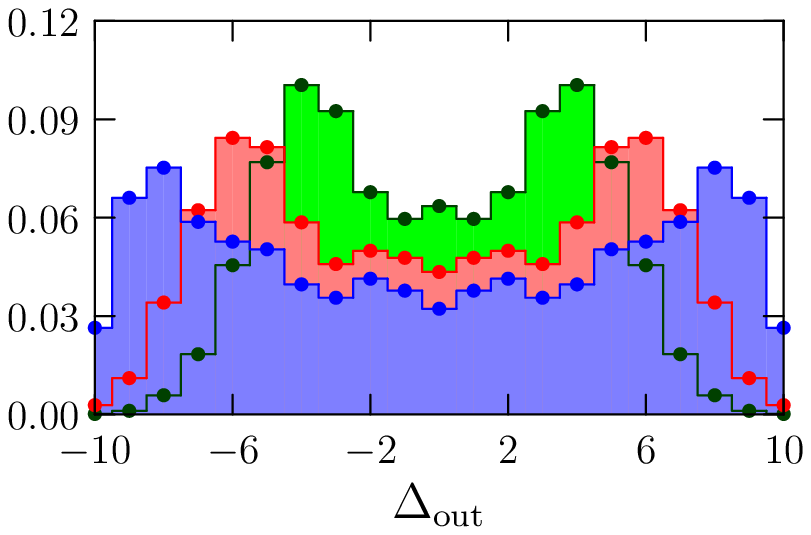}\\
\raisebox{2cm}{b)}&
\raisebox{1.4cm}{$0\%$}&
\includegraphics[height=2cm]{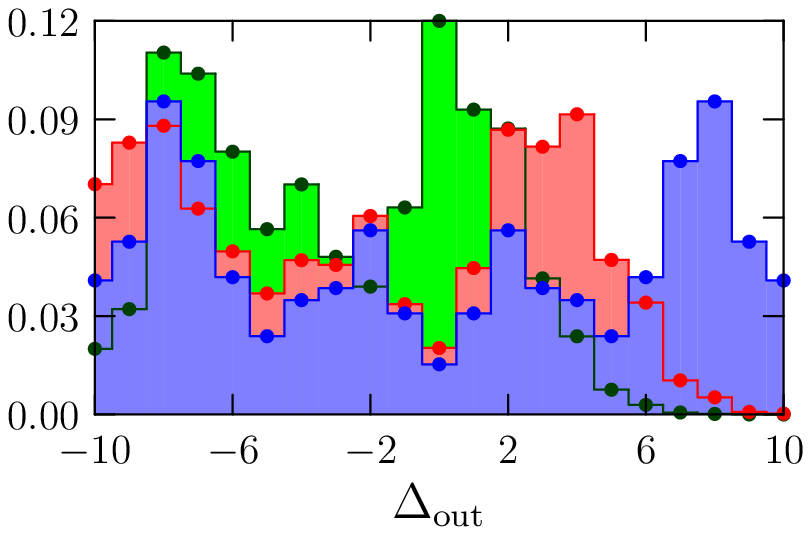}&
\includegraphics[height=2cm]{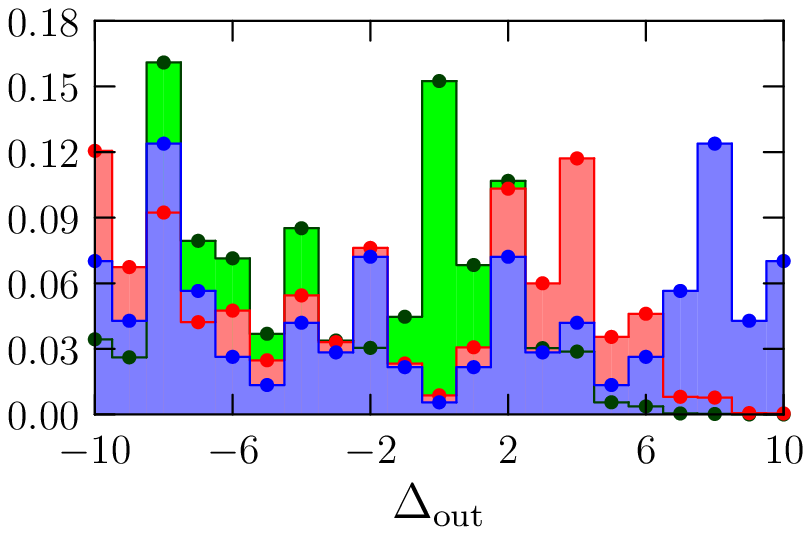}&
\includegraphics[height=2cm]{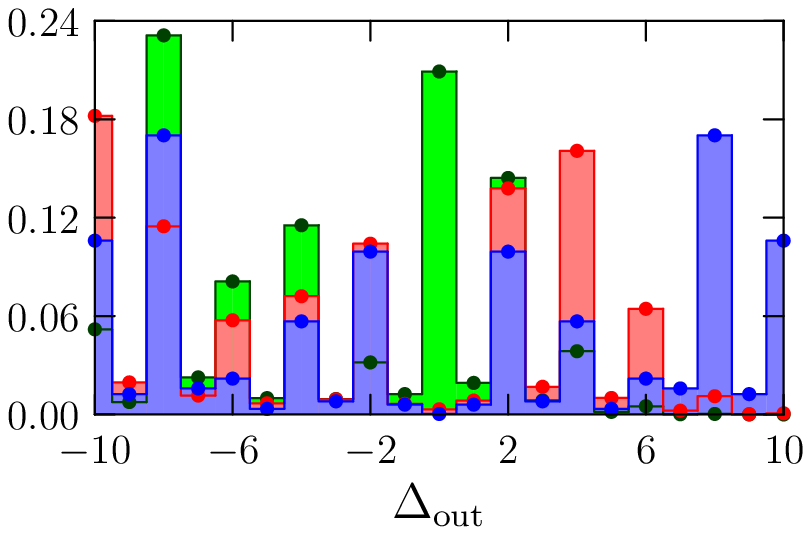}&
\includegraphics[height=2cm]{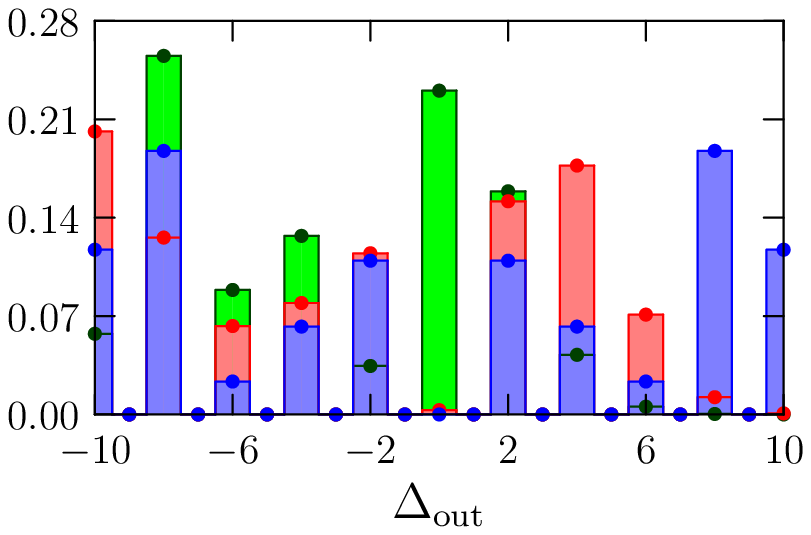}\\
&\raisebox{1.4cm}{$10\%$}&
\includegraphics[height=2cm]{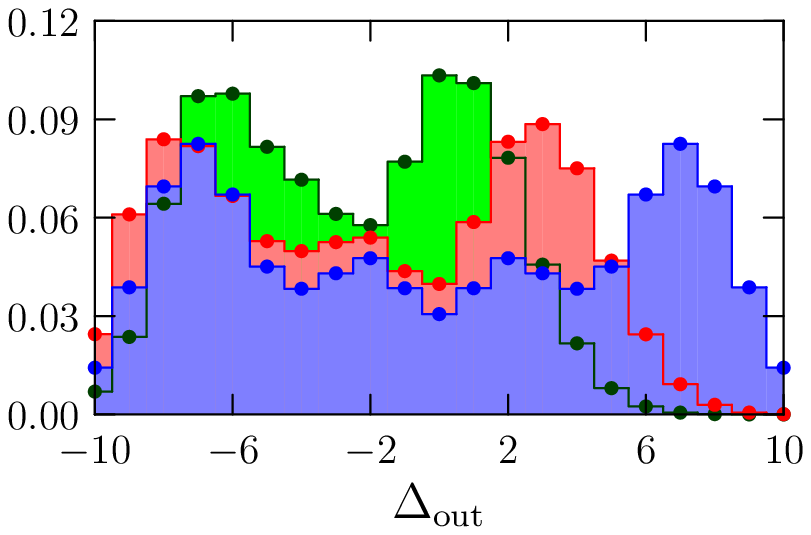}&
\includegraphics[height=2cm]{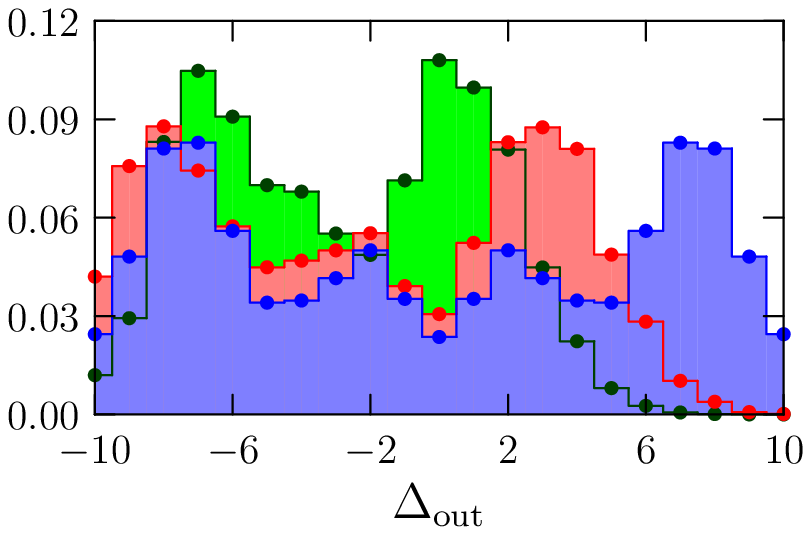}&
\includegraphics[height=2cm]{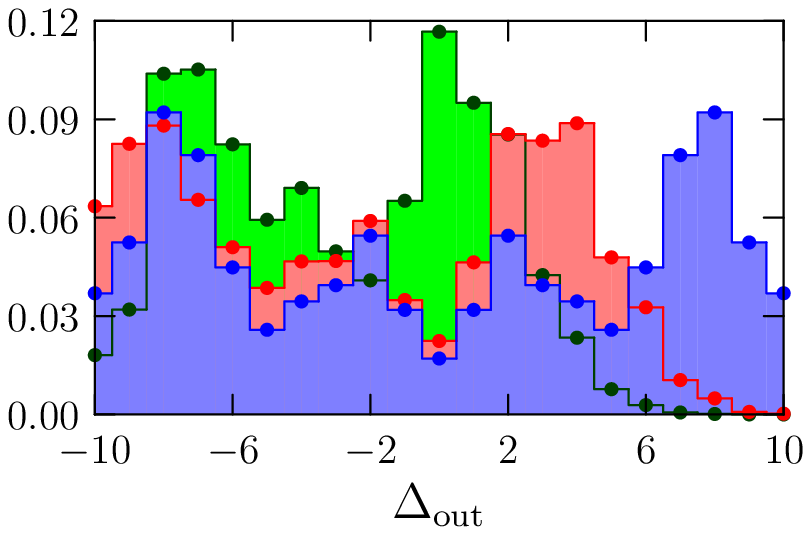}&
\includegraphics[height=2cm]{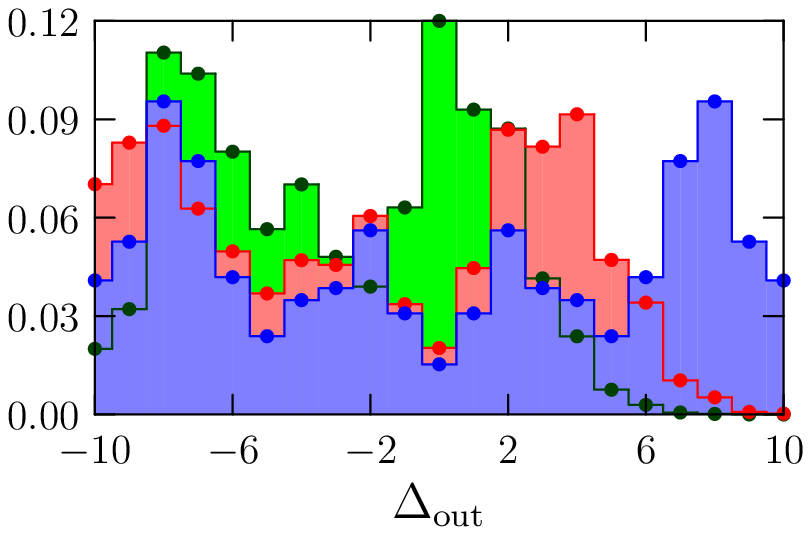}\\
&\raisebox{1.4cm}{$20\%$}&
\includegraphics[height=2cm]{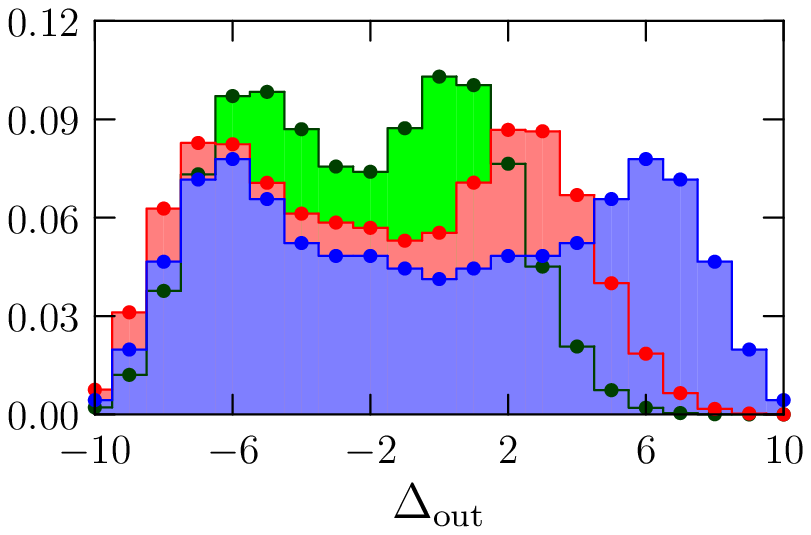}&
\includegraphics[height=2cm]{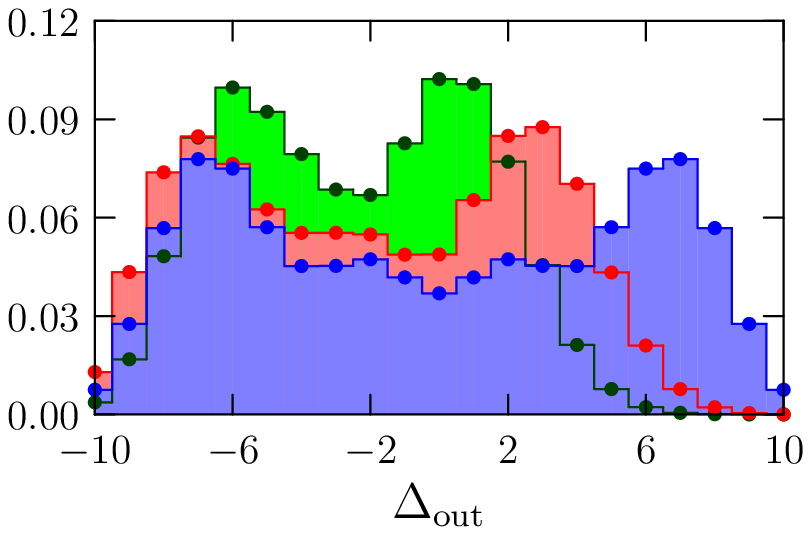}&
\includegraphics[height=2cm]{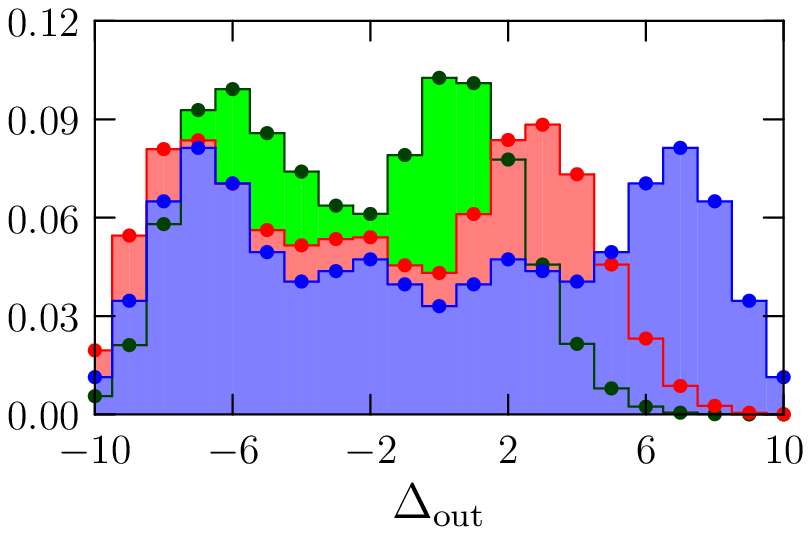}&
\includegraphics[height=2cm]{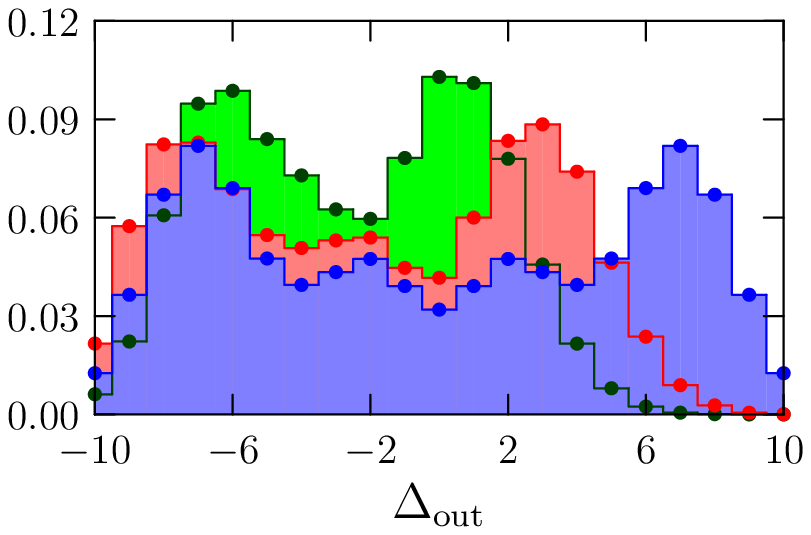}\\
\raisebox{2cm}{c)}&
\raisebox{1.4cm}{$0\%$}&
\includegraphics[height=2cm]{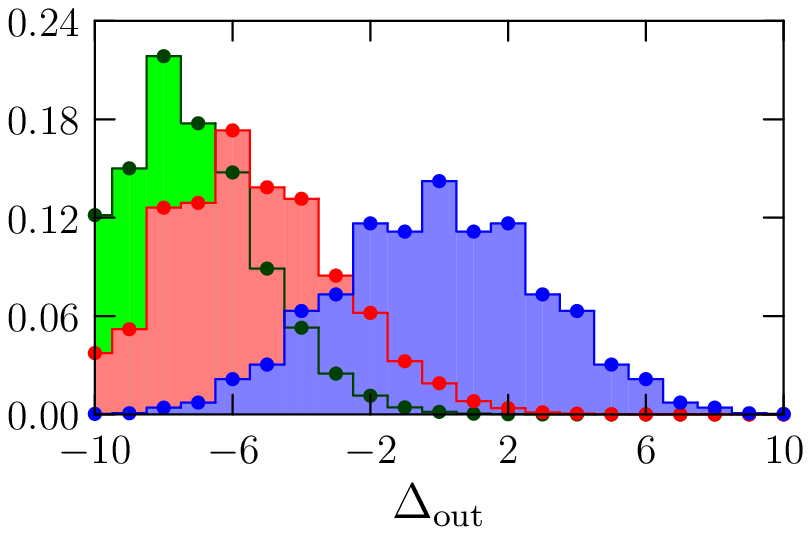}&
\includegraphics[height=2cm]{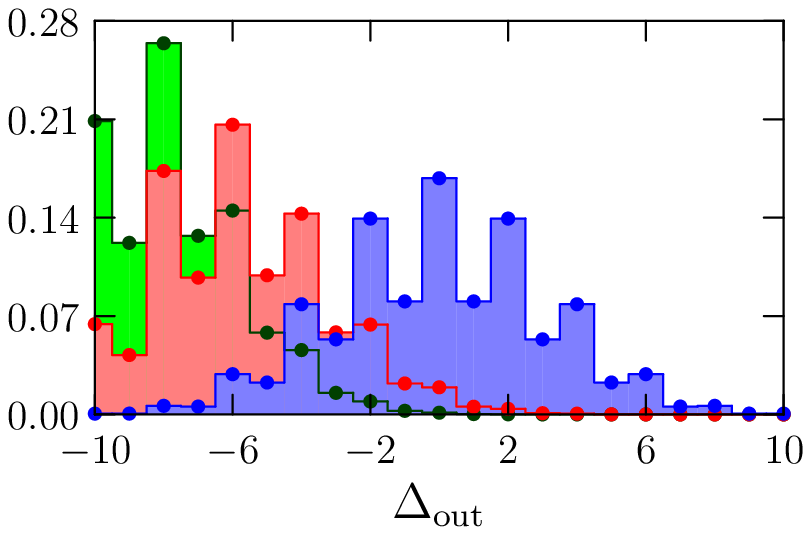}&
\includegraphics[height=2cm]{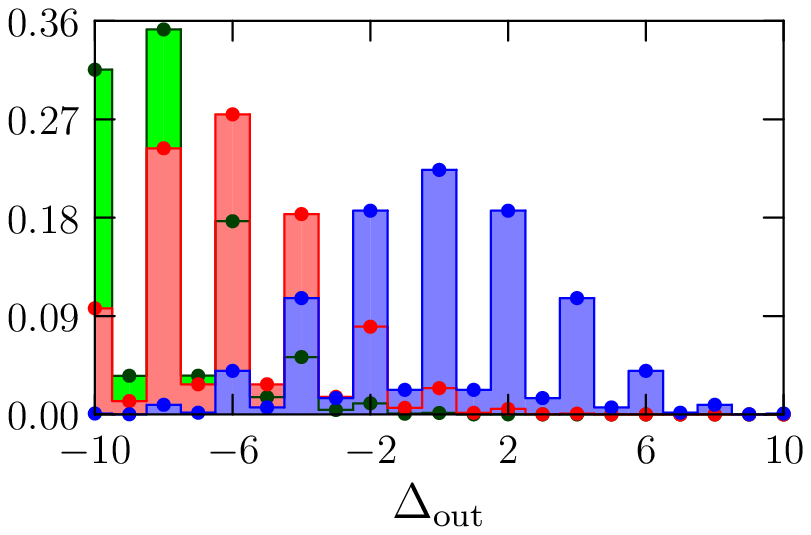}&
\includegraphics[height=2cm]{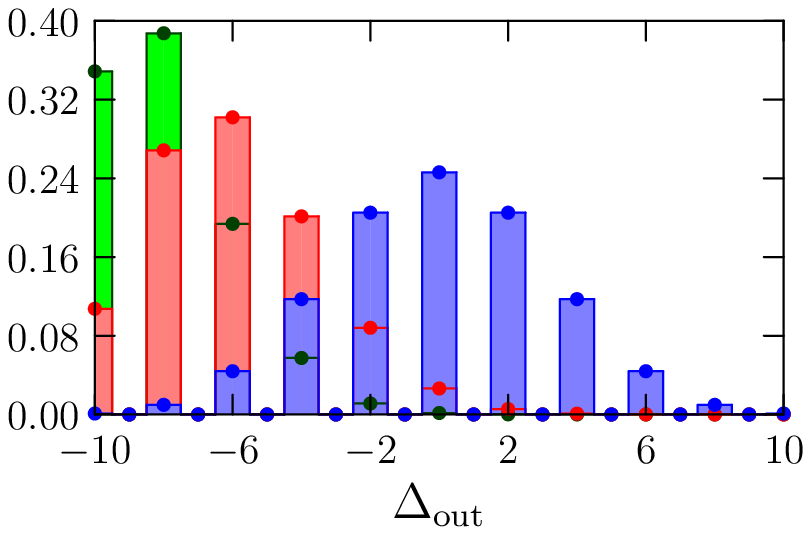}\\
&\raisebox{1.4cm}{$10\%$}&
\includegraphics[height=2cm]{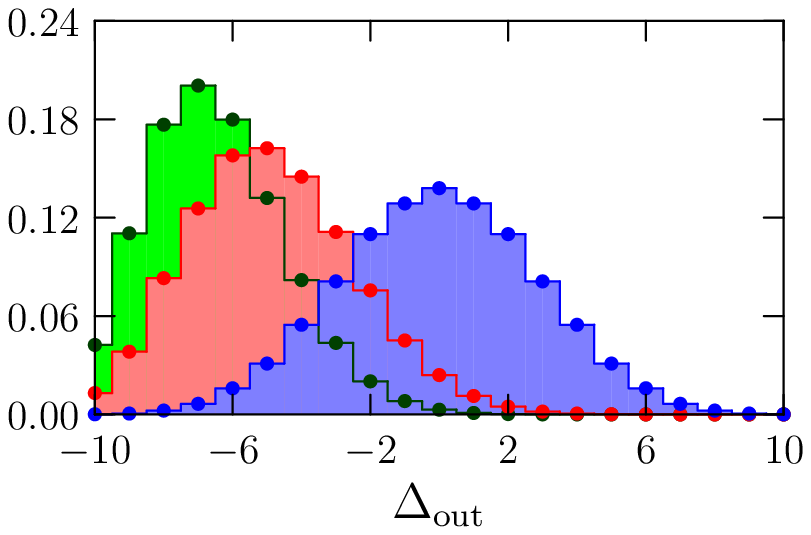}&
\includegraphics[height=2cm]{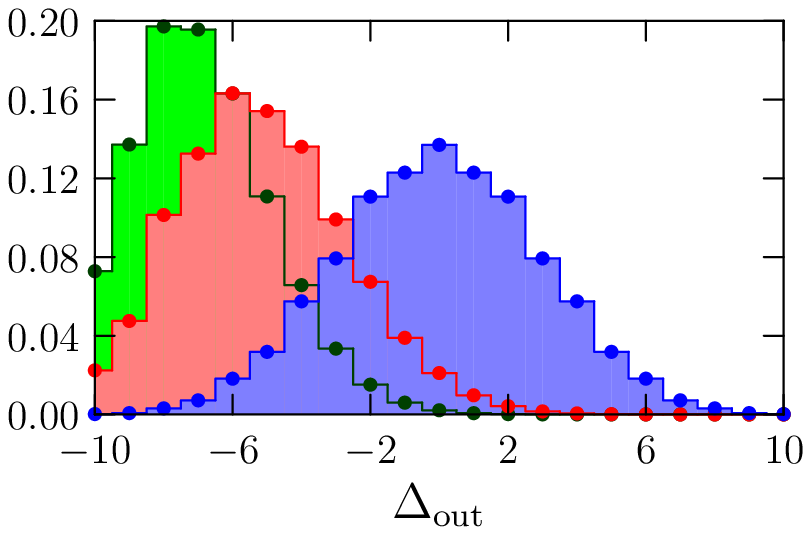}&
\includegraphics[height=2cm]{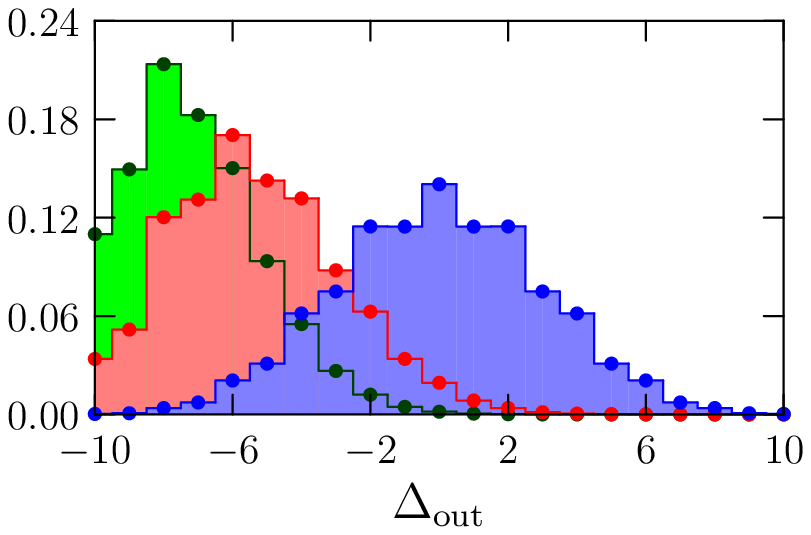}&
\includegraphics[height=2cm]{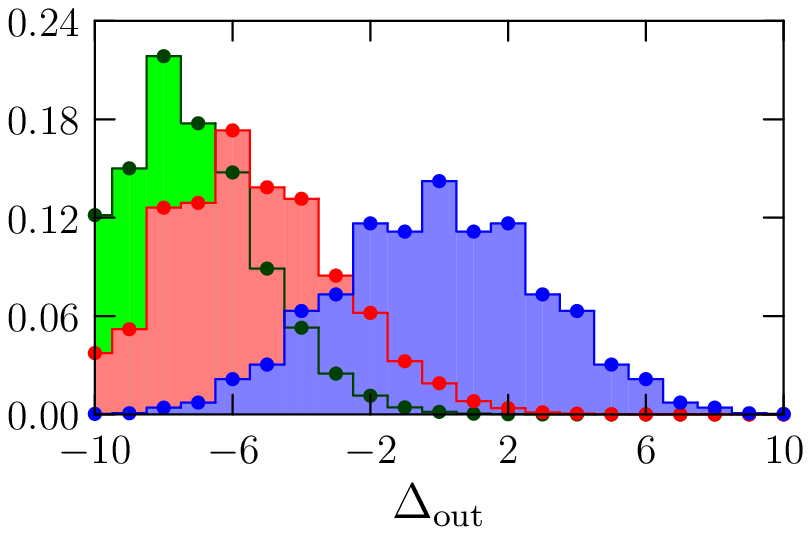}\\
&\raisebox{1.4cm}{$20\%$}&
\includegraphics[height=2cm]{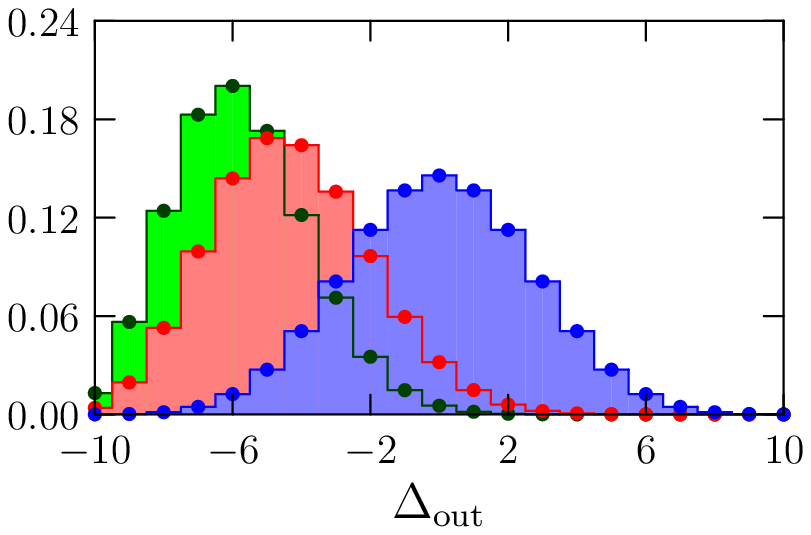}&
\includegraphics[height=2cm]{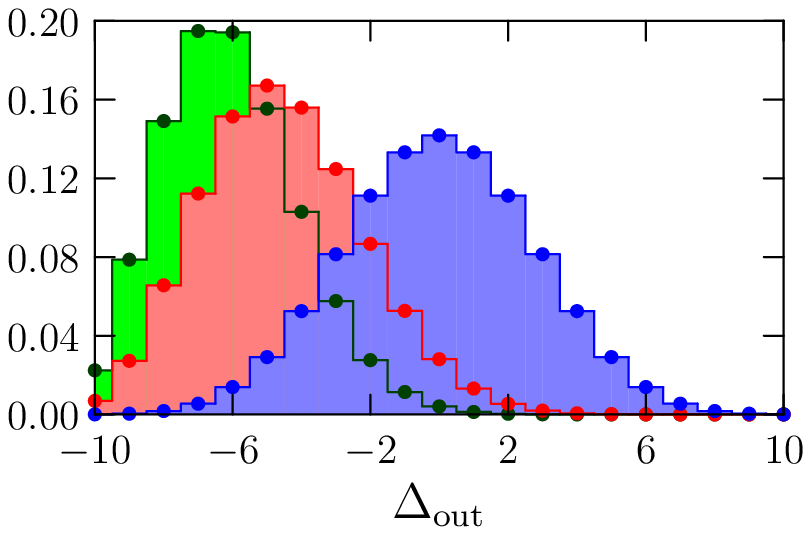}&
\includegraphics[height=2cm]{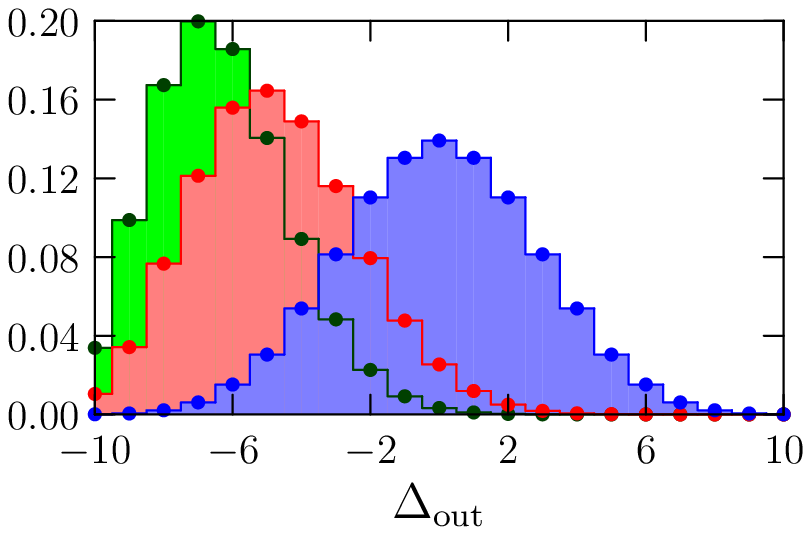}&
\includegraphics[height=2cm]{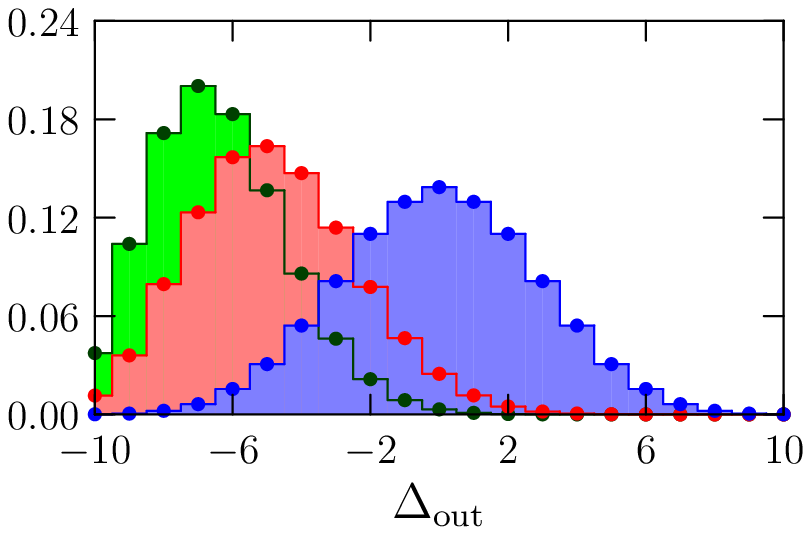}
\end{tabular}
\caption{(Color online) Probability distributions $p(\Delta_{\text{out}})$ from Fig.~\ref{fig:unbalancedQWS10} ($r =0.1$ -- green, $r =0.2$ -- red, $r=0.5$ -- blue) computed for various purities of input Fock states and amount of losses. $S=10$ a) $\Delta=0$, b) $-4$, c) $-10$.}
\label{fig:losses}
\end{figure*}

\end{document}